\documentclass[aps,letterpaper,onecolumn,superscriptaddress,prx,nofootinbib,floatfix]{revtex4-2}
\pdfoutput=1
\usepackage{amsfonts}
\usepackage{amssymb}
\usepackage[scanall]{psfrag}
\usepackage{psfrag}
\usepackage{graphicx}
\usepackage{xcolor}
\usepackage{braket}
\usepackage{tabularx}
\usepackage{amsmath,hyperref}
\hypersetup{colorlinks=true, linkcolor=blue!80!black, urlcolor=blue!80!black, citecolor=blue!80!black}

\begin{document}
\newcommand{\of}[1]{\left( #1 \right)}
\newcommand{\sqof}[1]{\left[ #1 \right]}
\newcommand{\avg}[1]{\left< #1 \right>}
\newcommand{\cuof}[1]{\left \{ #1 \right \} }
\newcommand{\pil}{\frac{\pi}{L}}
\newcommand{\bx}{\mathbf{x}}
\newcommand{\by}{\mathbf{y}}
\newcommand{\bk}{\mathbf{k}}
\newcommand{\bp}{\mathbf{p}}
\newcommand{\bl}{\mathbf{l}}
\newcommand{\bq}{\mathbf{q}}
\newcommand{\bs}{\mathbf{s}}
\newcommand{\psibar}{\overline{\psi}}
\newcommand{\svec}{\overrightarrow{\sigma}}
\newcommand{\dvec}{\overrightarrow{\partial}}
\newcommand{\bA}{\mathbf{A}}
\newcommand{\bdelta}{\mathbf{\delta}}
\newcommand{\bK}{\mathbf{K}}
\newcommand{\bQ}{\mathbf{Q}}
\newcommand{\bG}{\mathbf{G}}
\newcommand{\bw}{\mathbf{w}}
\newcommand{\bL}{\mathbf{L}}
\newcommand{\ohat}{\widehat{O}}
\newcommand{\up}{\uparrow}
\newcommand{\down}{\downarrow}
\newcommand{\MM}{\mathcal{M}}
\newcommand{\MN}{\mathcal{N}}
\newcommand{\MR}{\mathcal{R}}
\newcommand{\MT}{\mathcal{T}}
\newcommand{\MP}{\mathcal{P}}
\newcommand{\tW}{\tilde{W}}
\newcommand{\tX}{\tilde{X}}
\newcommand{\tY}{\tilde{Y}}
\newcommand{\tZ}{\tilde{Z}}
\newcommand{\tOm}{\tilde{\Omega}}
\newcommand{\barA}{\bar{\alpha}}
\makeatletter
\renewcommand\l@subsection[2]{} %
\renewcommand\l@subsubsection[2]{} %
\makeatother

\title{Roadblocks and Opportunities in Quantum Algorithms -- Insights from the National Quantum Initiative Joint Algorithms Workshop, May 20--22, 2024}

\author{Eliot Kapit}
\affiliation{Atom Computing and Colorado School of Mines}

\author{Peter Love}
\affiliation{Tufts University}

\author{Jeffrey Larson}
\affiliation{Argonne National Laboratory}

\author{Andrew Sornborger}
\affiliation{Los Alamos National Laboratory}

\author{Eleanor Crane}
\affiliation{King's College London}

\author{Alexander Schuckert}
\affiliation{Joint Quantum Institute, NIST/UMD}

\author{Teague Tomesh}
\affiliation{Infleqtion}

\author{Frederic Chong}
\affiliation{University of Chicago}

\author{Sabre Kais}
\affiliation{North Carolina State University}

\maketitle

The National Quantum Initiative Joint Algorithms Workshop brought together researchers across academia, national laboratories, and industry to assess the current landscape of quantum algorithms and discuss roadblocks to progress. The workshop featured discussions on emerging algorithmic techniques, resource constraints in near-term hardware, and opportunities for co-design across software and systems. Presented here are seven topics from the workshop, each highlighting a critical challenge or promising opportunity discussed during the event. Together, they offer a snapshot of the field’s evolving priorities and a shared vision for what is needed to advance quantum computational capabilities.

\section{Approximation Hardness for Quantum Algorithms}
\subsection*{Eliot Kapit, Atom Computing, Boulder, CO, and Colorado School of Mines, Golden, CO (Superconducting Quantum Materials and Systems Center), ekapit@mines.edu}
\subsection{Problem Statement }

The ultimate power and utility of quantum computers for exact and approximate optimization of hard constraint satisfaction problems (CSPs) \cite{korte2011combinatorial} is one of the most important open questions in quantum computer science. CSPs are ubiquitous in computer science; and since these problems generally are in the complexity class NP, they are in the worst and often typical cases are exponentially difficult for all known classical methods. Within the space of hard CSPs, one can observe that the most promising targets for achieving practical quantum advantage are problem classes or instances that are approximation-hard in practice. Informally, a problem is approximation-hard if all known classical methods struggle to return solutions within a defined fraction of the global optimum $G$'s cost function value. For many real-world CSPs---with a canonical example being logistics and scheduling problems---the practical value of returned solutions is a continuous function of the optimized cost function, and classical heuristics often do a  good job of returning solutions close in cost to the global optimum configuration $G$, with a ``difficulty cliff" setting close to the cost of $G$.\footnote{The existence of these difficulty cliffs has deep roots in the physics of spin glasses, where the onset of classical solution difficulty coincides with the emergence of a ``clustering phase" \cite{mezard2002analytic,mezard2005clustering,hartmann2006phase,krzakala2007landscape,altarelli2008relationship,gamarnik2021overlap} of many well-separated deep local minima. Finding any one of the exponentially many clusters is easy, but doing so gives no meaningful information about the existence of other, deeper solutions elsewhere in configuration space, making it extremely hard to find the optimal configuration. But in many problem formulations, the typical cluster energy is not far above that of $G$, and the difficulty cliff corresponds to an exponential collapse in the density of states as the energy of $G$ is approached.} If indeed the empirical location of this cliff is close to the cost of $G$, it may be counterproductive to try to overcome it; one can easily  imagine a hypothetical problem where one can get within 99\% of the global optimum on a laptop in a few minutes, but it takes weeks of supercompute (or quantum compute) time to get that last 1\%. For a real-world scheduling problem such as planning a delivery route,  a 1\% efficiency improvement is unlikely to be worth the enormous additional cost to find it.

Therefore, achieving useful quantum advantage for combinatorial optimization problems likely requires that these problems are approximation-hard in practice---or, in our language, the difficulty cliff sits far above the cost of $G$. This condition makes returned solutions more valuable. It also potentially makes them easier to find, by adding additional structure to the problem and, as argued in~\cite{kapit2024approximability}, by ruling out known hardness mechanisms in exact optimization.

\subsection{Previous Work}
Recently, a handful of papers have demonstrated beyond-quadratic speedups for quantum optimization, and it is not a coincidence that these works focus on approximation-hard problems, as well as requiring high circuit depths and fault tolerance. These include the novel method of decoded quantum interferometry for a range of problems \cite{jordan2024optimization}, Kikuchi matrices for the spiked tensor problem and planted noisy MAX-$k$-XORSAT \cite{hastings2020classical,schmidhuber2024quartic}, iterative (but non-variational) update methods such as IST-SAT \cite{barton2024iterative}, and spectrally filtered quantum optimization~\cite{kapit2024approximability}, discussed in more detail below. Critically, all these methods exploit structural features of the problem class, needed in order to obtain beyond-quadratic speedups given the optimality of Grover's algorithm \cite{zalka1999} and the widely believed Aaronson--Ambainis conjecture \cite{aaronson2009need} on the need for structure. In general, the more structure a problem has, the easier it is to solve with quantum methods, as exemplified in the recent work of Montanaro and Zhou \cite{montanaro2024quantum}, who showed that fully permutation-symmetric CSP instances are easy for low-depth QAOA, although they are  easy to solve classically as well. Furthermore, we note that the PCP theorem \cite{arora2009computational} implies that all hard CSP classes must also be approximation-hard in the worst case; the converse would imply P=NP. In many cases, including $k$-SAT and MAX-$k$-XORSAT, classically one cannot guarantee an approximation better than random guessing in polynomial time unless P=NP!

\begin{figure}
\includegraphics[width=0.49\columnwidth]{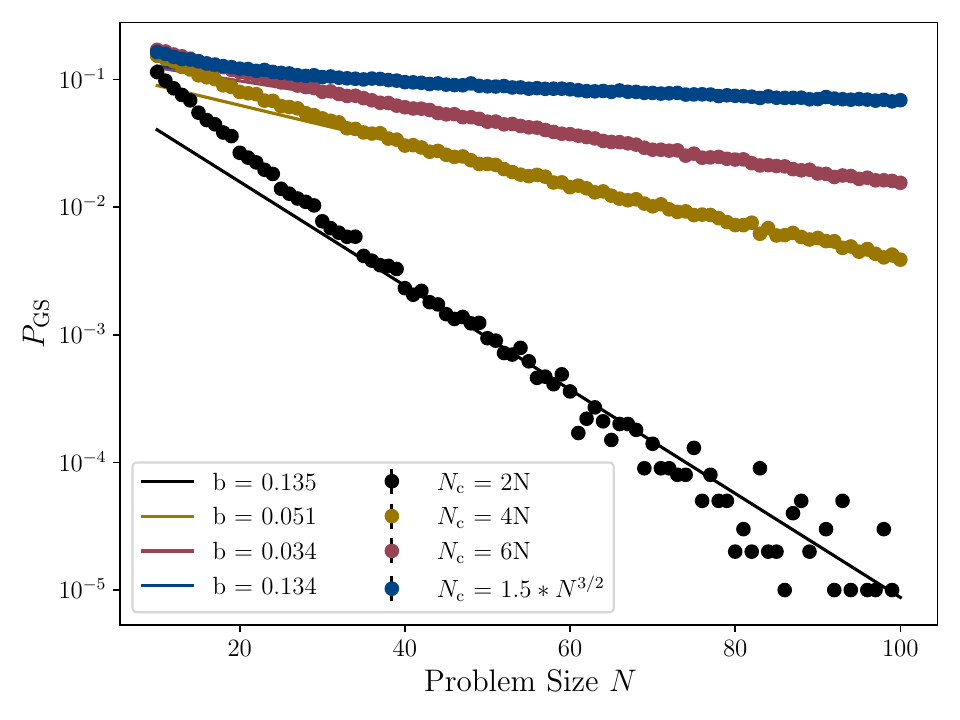}
\includegraphics[width=0.49\columnwidth]{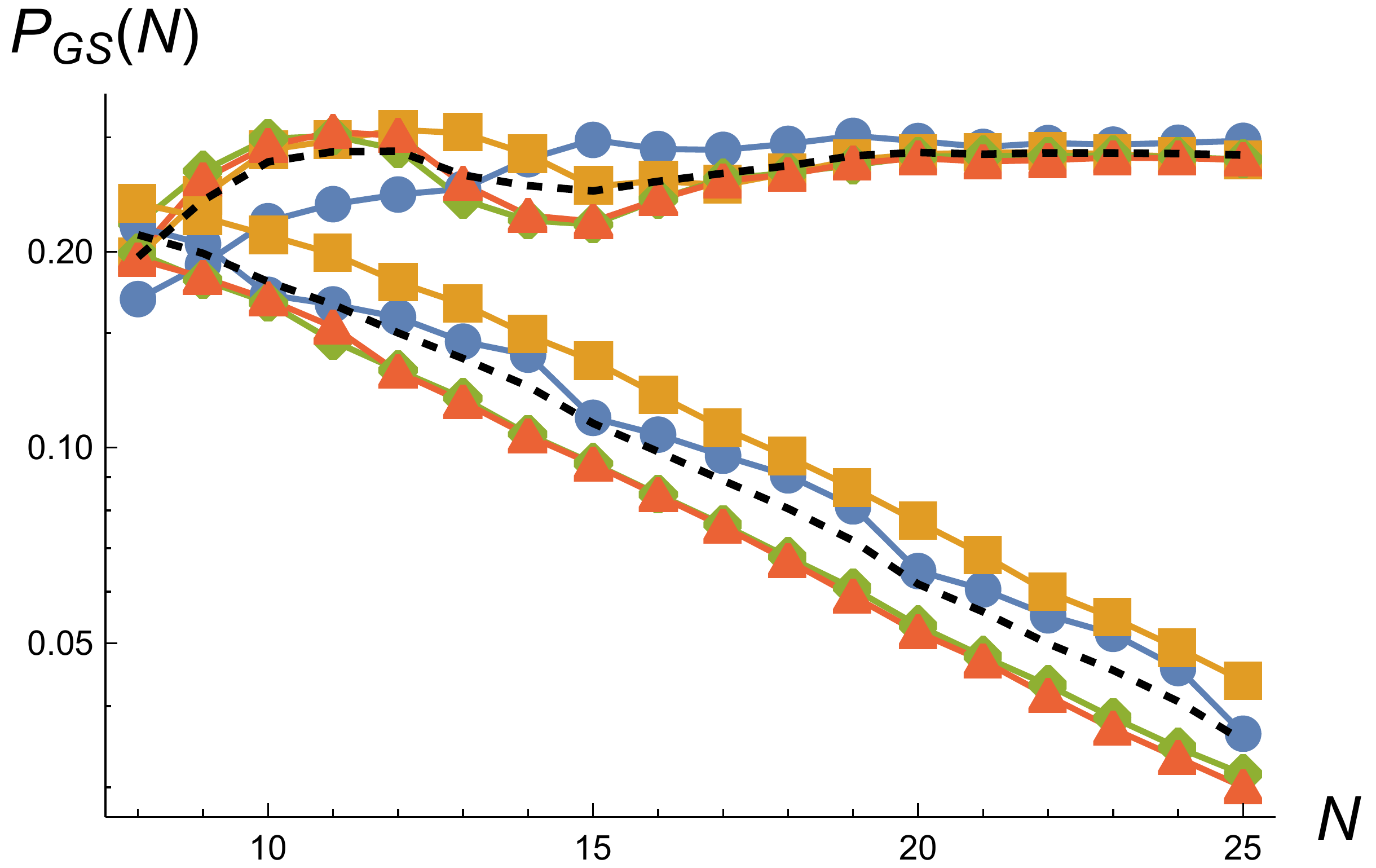}
\caption{Comparison of classical and quantum optimization heuristics, taken from~\cite{kapit2024approximability}, illustrating the central puzzle we describe here. Both plots display the per-shot success probability for finding the ground state of a set of planted noisy MAX-3-XORSAT instances vs. number of binary variables $N$; on the left for a classical quasi-greedy solver (with $O \of{N}$ per-shot runtime) and on the right for a pair of quantum routines, with planted unsatisfied fraction $\epsilon=0.1$ and constraint densities $d_C = \cuof{2,4,6,1.5 \sqrt{N}}$. The time to solution for classical methods is  sensitive to constraint density---essentially, due to entropic barrier physics~\cite{bellitti2021entropic}, these problems become easier as they get more extremal, although the time to solution remains superpolynomial. In contrast, for quantum methods, we observed almost no dependence on $d_C$. The bottom set of curves, for Trotterized AQC with $O \of{N^{3/2}}$ layers, all scale exponentially with the same exponent; and the top set of curves, for a spectrally filtered quantum optimization method at the same depth, converge to the same, nearly constant success probability. These results suggest that quantum and classical algorithms ``see" disorder and the cost function landscape qualitatively differently, and understanding this difference in more detail is a critical research goal moving forward.}\label{MSFOfig}
\end{figure}

With this in mind, we now ask: How powerful can fault-tolerant quantum computers be for approximation-hard instances? The answer is not clear; and perhaps surprisingly, whatever fundamental mechanisms would ensure approximation hardness for quantum methods are currently opaque. For classical algorithms based on local updates, ranging from simple greedy methods to message passing and parallel tempering, the intuitive mechanism for both exact and approximate optimization hardness is well understood: a phenomenon known as entropic barriers \cite{bellitti2021entropic}. Essentially, the entropic barrier concept captures the idea that unless one begins with a sufficiently lucky initial guess and starts near a sufficiently high-quality solution (an exponentially small fraction of the search space), optimization methods will instead converge to local minima potentially far from $G$ in both value and Hamming distance. The quantum analogue of this physics, however, is much less obvious, as illustrated in Figure~\ref{MSFOfig}. This problem was considered by one of us (and collaborators) for the prototypically hard MAX-3-XORSAT class in~\cite{kapit2024approximability}. In that work we argued that the known mechanisms ensuring exact solution hardness do not obviously ensure approximation hardness, and we proposed a new family of quantum algorithms called spectrally filtered quantum optimization (SFQO) aimed at resolving identified failure modes for QAOA \cite{farhi2014quantum} and adiabatic quantum computing \cite{albashlidar2017}. As seen in Figure~\ref{MSFOfig}, SFQO provides an apparent exponential speedup for solving approximation-hard instances, predicted through analytical (if approximate) theory and supported by numerical simulation. SFQO achieves its speedup by exploiting global statistical properties of these cost functions, globally applying nonlinear transformations to the cost function to amplify these correlations and find the solution much more quickly than QAOA and AQC.

Unless NP$\in$BQP---an unlikely possibility, to be sure---some set of physical processes must conspire to ensure approximation hardness at large scales. In the case of SFQO, either some as-yet-unknown mechanism must cause its time to solution to scale exponentially at much larger $N$, or some subset of approximation-hard instances must break its fundamental assumptions and thus ensure exponential scaling. Either way,  the impact of disorder and cost function landscape roughness for SFQO and other quantum methods is very different from that of their classical counterparts. This physics and the broader space of speedup mechanisms for optimizing CSPs with fault-tolerant machines are at best poorly understood. Understanding even some of the fundamental physical mechanisms at play here would lead to valuable discoveries in many-body physics and shed light on the ultimate power of quantum computing for this task. After all, sufficiently high-degree polynomial speedups can be  useful in practice \cite{babbush2021focus}. We thus conclude that a research program aimed at exploring novel, fault-tolerant quantum speedup mechanisms for approximation-hard CSPs and determining asymptotic failure mechanisms for real problem classes would be a valuable endeavor as fault-tolerant quantum computers become a practical reality over the next few years.

\section{Hybrid Quantum-Classical Algorithms and Applications: A Case Study
}

\subsection*{Teague Tomesh, Infleqtion, Chicago, IL, teague.tomesh@infleqtion.com \\ Frederic Chong, University of Chicago, Chicago, IL,  chong@cs.uchicago.edu}

\subsection{Multimodal Case Study}
Recent advances in quantum computing have generated significant excitement in the scientific community. Identifying practical applications that demonstrate clear quantum advantages, however, remains a persistent challenge.
While  multiple quantum algorithms  possess provable or suspected computational advantages, quantum computers will likely serve as domain-specific accelerators within larger hybrid quantum-classical architectures, rather than as standalone systems for end-to-end computation.
This case study examines our experience developing such a hybrid system for biomarker discovery in multimodal cancer data analysis, where quantum computing's potential as a specialized accelerator can be effectively leveraged.

The analysis of cancer data presents unique computational challenges that make it an interesting target for quantum-classical hybrid approaches \cite{ramesh2024quantum}. Cancer datasets typically feature high-dimensional data spaces, with each patient record containing thousands of features spanning genomic, transcriptomic, and imaging modalities \cite{echle2021deep}. The number of patient samples is often limited, however, because of the time and expense of data collection \cite{howard2021impact}. This imbalance between feature dimensionality and sample size can lead to overparameterization and overfitting in traditional machine learning approaches, necessitating more advanced feature selection methods \cite{echle2021deep, howard2021impact, fan_2014, hrinivich_2023}.

A particularly challenging aspect of feature selection in this domain is the desire to explicitly account for complex interactions between features in multimodal data.
The total number of these relationships grows combinatorially, presenting a daunting search space to sift through.
Yet these higher-order relationships often capture important biological mechanisms and can be crucial for accurate biomarker discovery \cite{watkinson2008identification, watkinson2009inference, boehm2022harnessing}. Furthermore, the medical nature of the application demands that our selected features remain interpretable to clinicians, ruling out various dimensionality reduction techniques that transform the original feature space such as principal component analysis or neural network-based methods such as sparse autoencoders.

Our approach aims to address these challenges by leveraging quantum algorithms, but the hybrid application must be carefully designed to distribute the workload between classical and limited quantum resources. We developed a hybrid algorithm (PCBO-Tournament) that frames feature selection as a polynomial constrained binary optimization (PCBO) problem and uses classical resources to break the single, large feature selection problem into smaller chunks that can be handled by the quantum computer. The algorithm leverages quantum resources through the Recursive Quantum Approximate Optimization Algorithm (RQAOA) \cite{bravyi2020obstacles} to efficiently explore higher-order feature correlations.

We highlight two primary roadblocks encountered during our algorithm development, describe our strategies for addressing them, and discuss promising directions for future quantum algorithms and applications research. First, proving quantum advantages for optimization applications remains challenging, leading us to emphasize practical implementation metrics over theoretical speedup guarantees and highlighting the need for robust benchmarking methodologies. Second, the severe limitations of current quantum computational resources---a constraint likely to persist into the foreseeable future---motivated our adoption of a co-design approach that carefully allocates quantum resources to specific portions of the computational workload where they can provide the most benefit.

\subsection{Roadblock -- Quantum Advantage for Optimization Applications}

Proving robust separations between quantum and classical algorithms for optimization problems is a fundamental challenge in quantum computing today. While some examples do exist---including the quadratic advantage of Grover's search algorithm \cite{grover1996fast}, upper bounds on the approximation ratio attainable by constant-depth QAOA \cite{bravyi2020obstacles, farhi2014quantum}, and exponential advantage in query complexity for random oracles \cite{yamakawa2024verifiable}---proving definitive separations for many practical applications remains elusive. This challenge is particularly acute for quantum optimization algorithms such as QAOA, which are primarily based on heuristics and can therefore resist theoretical analysis in higher-depth regimes.

However, the difficulty of proving theoretical asymptotic advantages need not diminish the potential practical value of quantum optimization algorithms. Consider, for example, satisfiability problems, such as the NP-complete 3-SAT problem, for which we have no known polynomial-time algorithm in the worst case \cite{karp1975computational}. Nevertheless, heuristic solvers for 3-SAT and related problems are widely deployed in industry, solving economically valuable problems in areas such as circuit design and automated reasoning \cite{gomes2008satisfiability}. This precedent suggests that studying and developing heuristic quantum optimization algorithms remains a valuable pursuit, even in the absence of provable advantages, and requires careful benchmarking to understand how these algorithms perform relative to existing approaches \cite{tomesh2022supermarq, lubinski2023application, lubinski2024optimization}.

Performance comparisons between quantum and classical algorithms consider multiple metrics beyond runtime, including solution quality and energy efficiency. Different use cases dictate which of these metrics matter most \cite{tomesh2022supermarq}.
In our work with feature selection for multimodal cancer datasets, solution quality typically takes precedence over runtime considerations. Given the precious nature of the data and the relatively small number of required algorithm runs, we can justify longer computation times if they yield higher-quality feature sets---as measured by downstream machine learning model accuracy. Nevertheless, comprehensive benchmarking across all relevant metrics remains essential for maintaining transparency about algorithmic trade-offs and justifying computational costs.

Our approach to feature selection illustrates these considerations in practice. We formulated feature selection as a PCBO problem to explicitly capture higher-order correlations within the dataset. This formulation represents feature sets as binary strings, where bits with value 1 indicate included features, and the optimization objective is to minimize the energy of the PCBO cost function. Notably, the formulation process itself has proven valuable to our oncology collaborators because it requires processing the cancer dataset to compute information-theoretic quantities such as mutual and interaction information and piecing these together into a cost function such that low-energy solutions correspond to high-accuracy feature sets.
The resulting optimization problem takes the form of a multispin glass with all-to-all connectivity.  Finding low-energy solutions to such problems challenges both quantum and classical computers, and the best available tools remain heuristic  \cite{yasuoka2021computational}. In this context, fair comparison between classical, quantum, or hybrid solvers requires rigorous head-to-head benchmarking under carefully controlled conditions.

Multispin problems like these represent particularly promising candidates for potential quantum advantages in optimization. Classical algorithms begin to struggle with these problems at relatively modest scales---on the order of hundreds of binary variables, corresponding to the number of qubits required for quantum implementation. This aligns with prior work studying the potential for quantum advantage in optimization, which identified similar challenging problems such as low autocorrelation binary sequences \cite{shaydulin2024evidence} and multidimensional knapsack problems \cite{abbas2024challenges}. Quantum circuits can straightforwardly represent the higher-order correlations in the PCBO problems through CNOT ladder constructions, although doing so introduces significant demands on hardware resources, particularly regarding parallelism, circuit depth, and qubit connectivity.

These hardware considerations lead us to our next major challenge: designing quantum optimization algorithms that can execute effectively on quantum processors while maintaining their theoretical advantages. As we  discuss in the following section, this requires careful attention to hardware constraints and innovative approaches to algorithm design.

\subsection{Roadblock -- Limited Quantum Computational Resources}

\subsubsection{Quantum I/O Constraints}

A fundamental challenge in developing quantum applications for classical data analysis is managing the severe limitations on data input and output  imposed by quantum mechanics, particularly Holevo's bound \cite{holevo1973bounds}. These constraints force us to carefully consider how classical data is encoded into quantum states and subsequently measured. As an example application, feature selection presents itself as a use case that aligns naturally with these quantum I/O restrictions.

The appealing aspect of a feature selection application is its low overhead in terms of data representation requirements. Rather than encoding actual feature values into quantum states---which can be prohibitively expensive for certain data types, for example when representing continuous real values to floating-point precision--- a qubit need only represent whether a feature is selected (1) or not (0). The actual feature values appear in the classically preprocessed edge weights in the PCBO problem graph, enabling us to work with arbitrary classical data types while maintaining a consistent quantum representation. This approach incurs a linear scaling relationship between the number of data features and required qubits, making it feasible to process real-world cancer datasets with future quantum computers with thousands of logical qubits \cite{howard2021impact, hieromnimon2023latent}.

By considering the I/O constraints of quantum processors, we can observe a general pattern for ideal quantum applications: compact input representation, exponential search space during computation, and minimal output requirements. For feature selection, the problem can be initialized efficiently by creating an equal superposition over all possible feature combinations or by searching directly in a constrained subspace  for a particular number of features. When measuring the final state, we only need to extract a binary string indicating the selected features, naturally aligning with quantum measurement limitations.

\subsubsection{Slow and Imperfect Gate Operations}

A significant challenge in implementing quantum algorithms for practical applications lies in the fundamental limitations of quantum computational hardware, particularly regarding gate operations. While quantum computers offer theoretical advantages for certain computational tasks, their practical utility is constrained by both slow clock speeds and imperfect gate operations \cite{tomesh2022supermarq, lubinski2023application, lubinski2024optimization, babbush2021focus, hoefler2023disentangling}. These limitations impose practical caps on both the depth of circuits that can be reliably executed and the total number of circuits that can be run within reasonable timeframes. However, design choices at the application and compiler level can help  mitigate these hardware constraints.

For hybrid variational algorithms \cite{cerezo2021variational} such as the QAOA, the challenges of imperfect and slow logic operations are particularly apparent. Typically, variational algorithms require many circuit evaluations within an inner loop to measure the expectation value of the relevant cost function. This measurement is then used within an outer loop where a classical optimizer determines the parameter updates to begin the next iteration. Returning to the example of feature selection, we typically require fairly deep circuits, scaling with the order of the PCBO cost function, to generate high-quality solutions. When these circuit execution requirements are combined with slow gate operations, the wall-clock runtime of these applications quickly becomes prohibitive.

Recent work addresses these limitations of the QAOA by leveraging a parameter transfer approach that strategically balances classical and quantum resources \cite{galda_transferability_2021, shaydulin_parameter_2023, augustino2024strategies, hao2024end}. This method offloads the expensive variational optimization components to classical computers, requiring only a small number of deep quantum circuits to be executed on the actual quantum hardware. This approach can be viewed as an expensive but valuable compilation technique, particularly justified for NP-hard optimization problems where solution quality warrants the classical overhead. By simulating many smaller tractable problems classically to extract optimal parameters, we can then execute a minimal number of deep circuits on the quantum computer to obtain solutions at the desired problem size.

Complementary approaches to managing these hardware constraints include exploring space-time trade-offs on platforms such as neutral atoms, where qubits are relatively inexpensive compared with gates \cite{saffman2010quantum, manetsch2024tweezer}. By utilizing ancillary qubits combined with mid-circuit measurement and gate teleportation \cite{niu2023powerful, hashim2024efficient}, one can significantly reduce overall circuit depth. Another promising direction is the RQAOA, which executes multiple, shallower rounds of the QAOA, rather than a single round with a deeper ansatz. While this approach increases total runtime, it addresses a fundamental limitation of constant-depth QAOA, and it can achieve better solution quality by effectively growing circuit depth across multiple rounds, making it more amenable to quantum hardware constraints.

These approaches represent different strategies for navigating the practical limitations of quantum hardware while maintaining algorithm effectiveness. The choice between them depends on specific hardware capabilities, problem requirements, and the relative costs of classical versus quantum resources in the target system.

\subsection{Challenges and Opportunities for Cancer Resesarch and Other Domains}

The development of hybrid quantum-classical applications for cancer biomarker discovery illustrates both the challenges and opportunities in practical quantum computing. While proving theoretical quantum advantages remains difficult, especially for optimization problems, our experience demonstrates that focusing on practical metrics and careful co-design has the potential to yield meaningful improvements in solution quality. The constraints imposed by  quantum computational hardware---from Holevo's bound to imperfect gate operations---favor innovative algorithmic approaches that intelligently partition workloads between classical and quantum resources. Our feature selection case study exemplifies how quantum computing can be effectively leveraged even within these limitations, particularly when the application naturally aligns with quantum mechanical constraints. As quantum hardware capabilities continue to evolve, the lessons learned from this co-design approach will become increasingly valuable for developing the next generation of hybrid quantum-classical applications, not just in cancer research, but across a broad spectrum of computational challenges where domain-specific acceleration can provide tangible benefits.

\subsection*{Acknowledgments}
Work on this manuscript is supported by Wellcome Leap as part of the Quantum for Bio Program, as well as supported in part by by STAQ under award NSF Phy-1818914/232580; in part by the US Department of Energy Office of Advanced Scientific Computing Research, Accelerated
Research for Quantum Computing Program;  in part by the NSF Quantum Leap Challenge Institute for Hybrid Quantum Architectures and Networks (NSF Award 2016136); in part based upon work supported by the U.S. Department of Energy, Office of Science, National Quantum
Information Science Research Centers; and in part by the Army Research Office under Grant Number W911NF-23-1-0077.
This research used resources of the National Energy Research Scientific Computing Center, a DOE Office of Science User Facility using NERSC award NERSC DDR-ERCAP0030280.

\section{Simulating Fermions and Bosons}
\subsection*{Eleanor Crane, King's College London, Strand, London WC2R 2LS, eleanor.crane@kcl.ac.uk, \\ Alexander Schuckert, Joint Quantum Institute, NIST/UMD, College Park, MD,  aschu@umd.edu} %

Quantum simulation is the use of quantum devices for simulating quantum phenomena in nature, including materials science, chemistry, nuclear physics, and high-energy physics. In all of these fields, the fundamental task is to simulate interacting systems of bosons (such as phonons, photons, molecular vibrations and the Higgs particle) and fermions (such as electrons and protons). Important example problems are phonons coupling to electrons leading to superconductivity, or the down-conversion of light during photosynthesis. Simulating such problems could even have societal benefit, for instance for the design of lithium-ion batteries~\cite{Delgado_2022,ShokrianZini2023}. Fermion-boson problems are believed to be a generically classically hard problem, which is why simulating such systems is one of the most promising applications of quantum computers. In particular, large boson number fluctuations such as those appearing in bunching phases render even state-of-the-art methods such as matrix product states challenging~\cite{schuckert2025constrainedmanybodyphasesmathbbz2higgs}. However, despite digital qubit-based quantum computers being universal and therefore holding the promise of being able to solve these problems, notoriously large numbers of qubits and logic gates are required to represent fermion-boson problems. This is because of the theoretically infinite Hilbert space of the bosonic modes, which must be approximated with qubits. In addition, time-evolving under bosonic Hamiltonians comes with an overhead which is polynomial in the cutoff, with a large constant prefactor. For fermions, the anti-commutation of fermionic operators needs to be emulated by commuting qubit operators, which also leads to an at worst linear overhead in the number of two-qubit gates. Therefore, simulating fermions and bosons with qubit quantum computers is challenging and far harder than simulating the spin models which most current hardware demonstrations have focused on. In the following, we will discuss these roadblocks in more detail, and present several possible directions to overcome them. We first discuss bosons, then fermions.

\begin{figure}[h]
\centering
\small
    \def\arraystretch{1.5}
    \begin{tabular}{|c|c|c|c|c|}
    \hline
     & Algorithm Scaling Improvement & Definition of N \\ \hline
        Qubit-boson processor & $\tilde{\mathcal{O}}(N)$ gates for qubits $\rightarrow \mathcal{O}(1)$ ~\cite{crane2024hybrid} & Local Hilbert space dimension\\ \hline
        Qubit-fermion processor & $\mathcal{O}(N) $ depth for qubits $\rightarrow \mathcal{O}(\log_2(N))$ ~\cite{schuckert2024fermion} & Lattice size of crystalline material \\
        \hline
    \end{tabular}
\caption{Examples of speedups demonstrating the advantages of quantum computing using native matter particles for quantum simulation. Qubit-boson logarithmic speedup from Ref.~\cite{crane2024hybrid} for simulating a single bosonic hopping. The number of qubits representing a single bosonic site scales as $\mathcal{O}((\mathrm{log}_2(N))^2)$. Qubit-fermion speedup from Ref.~\cite{schuckert2024fermion} for simulating a single Trotter step under the materials Hamiltonian. Ref.~\cite{schuckert2024fermion} also showed a qubit algorithm with $\mathcal{O}(\log_2(N))$ depth, but with prohibitively expensive $\mathcal{O}(N^2)$ ancillas.  \label{fig_intro}}
\end{figure}

\subsection{Challenges for bosons}
In order to simulate bosonic systems with qubit quantum computers, we must first map the bosons to qubits and then simulate time-evolution under bosonic Hamiltonians, which requires mapping the bosonic Hamiltonian to a qubit Hamiltonian.

Bosons can be mapped to qubits by either sampling the phase space~\cite{Jordan_Lee_Preskill_2012} or representing the mode occupations in Fock space~\cite{Shaw2020, crane2024hybrid}. In both cases, it is beneficial to write the boson operator as a ``mean-field part'', either for the annihilation operator $\hat a = \braket{\hat a} +\delta \hat a$, or the number operator $\hat n = \braket{\hat n} +\delta \hat n$, for phase space, or Fock space encodings, respectively. Only the fluctuations $\delta \hat a$ and $\delta \hat n$ then need to be represented with qubits. From this discussion, it is clear that the regime which requires a large number of qubits is the one with large boson density fluctuations. Which of the two discretisations are beneficial then depends on whether the fluctuations are small in either phase or number space. Multiple ways exist to represent a Fock state with qubits, either by the unary or "one-hot" encoding (which requires one qubit per Fock state), or the Gray and binary encodings (which require only $\log_2(K)$ qubits for cutoff $K$).

The larger challenge of bosonic simulation is not the mapping of the state space, but the mapping of time evolution under bosonic Hamiltonians to single and two qubit gates. In Ref.~\cite{Sawaya2020} for example, different Fock-space encodings have been optimized and compared. In this estimate, simulating just a single beamsplitter with cutoff $K=7$ requires on the order of one to two thousand gates for the gray or binary encodings~\cite{Sawaya2020}. The unary or "one-hot" encodings are considered wasteful as they require as many qubits as states in the Hilbert space and do not lead to corresponding savings~\cite{Sawaya2020}. While this work focused on optimising low-$K$ simulation by numerical optimisation, a scalable (to large $K$) algorithm has been proposed in Ref.~\cite{crane2024hybrid, liu2024}. However, this leads to much larger gate estimates than those from numerical optimisation in the low-$K$ regime, showing that by no means, optimality for boson simulation has been reached.

At the heart of the challenge of simulating bosons with qubit hardware is the calculation of the square-roots issued from the action of the creation or annihilation operators, e.g. $\hat a \ket{n} = \sqrt{n} \ket{n-1}$, where $\hat a$ is the annihilation operator and $\ket{n}$ is the Fock state with $n$ bosons, and $n$ is an integer. Arithmetic is challenging for qubit quantum computers, so one option is to pre-calculate the square-roots which is made easy using the unary or "one-hot" encoding as each bit is assigned a value and the square-roots can therefore be applied bit-wise. However, as previously mentioned, these encodings are highly wasteful in terms of qubits and gates. Another possible way to calculate square-roots is using Newton iterations~\cite{häner2018optimizingquantumcircuitsarithmetic}, but it is extremely costly and approximate.

One way out of this problem is to use controllable bosonic degrees of freedom directly rather than mapping to qubits. This direction is motivated by recent impressive developments in the use of bosonic modes in circuit QED~\cite{Wang2023}, ion traps~\cite{whitlow2023} and neutral atoms~\cite{shaw2025}. Much of the progress has been done in the direction of bosonic error-correction, however the set of gates used in this context couples bosons to qubits. The qubits provide a non-linearity to the bosonic modes, enabling universal quantum computation. This is exactly what is required for quantum simulation of fermion-boson systems, mapping the fermions to qubits, and the bosons to native bosons in the hardware. This promising direction has been explored for quantum simulation in Ref.~\cite{crane2024hybrid}. A wrapper for Qiskit enabling initial circuit usage is described in~\cite{biskit} which can use compilation strategies devised in~\cite{Kang2025}. Initial steps to solving the compilation problem of compiling arbitrary bosonic operations onto single- and two-oscillator/qubit gates have been made in Ref.~\cite{decker2025,decker2025kernpilercompileroptimizationquantum,chen2025}. A big challenge is how to extend error-mitigation or error correction methods from oscillators into oscillators. While some error-correcting codes have been discussed \cite{Noh_o2o, xu2024lettingtigercagebosonic} and error-mitigation methods for analog simulators might be generalizable~\cite{steckmann2025errormitigationshottoshotfluctuations}, it is as of yet unclear how to perform universal computations on such logical bosonic modes.

\subsection{Challenges for fermions}

For fermions, there is no issue in the mapping of the Hilbert space: a fermionic mode can be either occupied or not, therefore directly mapping to the two qubit states. However, when time-evolving, the anti-commutation relations must be preserved, for which a mapping of fermionic operators to qubit operators is necessary. In the Jordan-Wigner encoding, weight-two fermion operators map to a weight-$N$ (for $N$ fermionic modes) qubit operator in the worst case, leading to a $\mathcal{O}(N)$ depth and gate overhead in the worst case. For local Hamiltonians, especially in 2D, more weight-efficient encodings are available, but require $\mathcal{O}(N)$ ancillas~\cite{Verstraete_2005, Derby_compact_2021, luo2025}, but even in that case,  a large constant (in $N$) overhead in terms of gates and depth makes simulation of fermionic models far more challenging than spin models. In fact, few demonstrations on  quantum hardware have been performed outside of the trivial $1$D case (where nearest-neighbour-interacting fermionic models map to nearest-neighbour-interacting spin models). As far as we are aware, the largest simulations to date are a $2\times 8$ lattice of spinful fermions ($32$ qubits in Jordan-Wigner encoding), up to two Trotter steps~\cite{hemery2024}, a $6\times 6$ lattice of spinless fermions~\cite{nigmatullin2025} up to four Trotter steps (using $36$ qubit in Jordan-Wigner and $48$ in the compact encoding), both in trapped ions, and a $4\times 4$ lattice of spinful fermions~\cite{evered2025} up to three Trotter steps in neutral atoms ($64$ qubits in a Honeycomb Majorana encoding, whose dynamics conserves the particle number only approximately). This is to be contrasted by the large-scale and long-time digital dynamical simulations that have been performed on spin models~\cite{kim2023,haghshenas2025} of $50-130$ spins up to dozens of Trotter steps.

Similarly to bosons, one way out of this problem is to use natively available fermionic degrees of freedom in quantum hardware, such as fermionic neutral atoms~\cite{yan2022, gonzalez-cuadra2023} for which the advantage over qubits was shown in Ref.~\cite{schuckert2024fermion} and electrons in quantum dots~\cite{wang2022a, rad2024analogquantumsimulatorquantum}.  This approach can be error corrected~\cite{schuckert2024fermion,ott2024}, and even made fault-tolerant~\cite{schuckert2024fermion}, offering large depth and gate advantages compared to qubits.

\subsection{Future Directions}

We propose several directions to overcome the difficulties mentioned above:
\begin{itemize}
    \item Develop quantum computing architectures including native fermions and bosons. Because of the large Hilbert space of bosonic modes, many of the compilation strategies developed in the context of qubit quantum computing do not apply. This means that in the context of the novel instruction set architectures offered by the hardware, new compilation strategies need to be developed both for quantum simulation and algorithms, but also error-correction.
    \item Identify which practically relevant problems actually require the simulation of quantum bosonic degrees of freedom. For instance, to first order, nuclear dynamics can be simulated classically in chemistry and therefore do not require quantum bosonic simulation. However, one candidate problem is the simulation of toy models of photosynthesis in which vibrations are quantum dynamical~\cite{OReilly2014}. Furthermore, the simulation of classical continuous variables could be sped up using quantum continuous variable, for instance to model dissipative nonlinear differential equations relevant to fusion~\cite{Liu2021}. This will require inter-disciplinary research discussions with experts from the target applications, especially those without exposure to quantum computing.
    \item Develop heuristic hybrid quantum-classical algorithms. For instance, the variational quantum eigensolver (VQE) does not necessarily need to employ unitaries connected to the Hamiltonian, which means that boson/fermion-to-qubit mappings might be unnecessary. One question is therefore how far one can push ansätze which are derived from qubit Hamiltonians for describing fermion and  bosonic systems. Another direction is quantum-classical Monte Carlo for finite-temperature properties~\cite{lu2021,schuckert_2023,ghanem2023}, which only requires short-time evolution and therefore the fermion-/boson-to-qubit mapping is not as important.
    \item Use of measurements, i.e. local operations and classical communication, to propagate information about fermionic statistics and large bosonic Hilbert spaces. A first step into this direction is the preparation of spin-$1$ states on qubits using measurement and feed forward~\cite{PRXQuantum.4.020315}.
\end{itemize}

\section{Quantum Machine Learning}
\subsection*{Sabre Kais, Department of Electrical and Computer Engineering, North Carolina State University, Raleigh, NC, skais@ncsu.edu}
\subsection{Neural Network Quantum States  Training with Quantum Devices}
The development of a \textit{polynomially efficient quantum-enabled variational Monte Carlo scheme} \cite{sajjan2024} represents a transformative advancement in leveraging quantum algorithms for training neural network quantum states (NQS). Renowned for their extensive expressibility, NQS provides a compelling alternative to traditional variational \textit{ansätze}. However, classical training methods often face significant computational bottlenecks, including slow convergence and challenges in representing high-dimensional quantum states with intricate correlations.

Our proposed method   overcomes the challenges of classical NQS training by employing quantum-enhanced Monte Carlo techniques to achieve efficient sampling and faster convergence \cite{sajjan2024}. By leveraging the linear scaling of circuit width and depth, the approach ensures compatibility with near-term quantum devices while maintaining the flexibility to explore complex quantum state spaces. These features enable the algorithm to accurately capture intricate correlations and extend the applicability of NQS to systems previously inaccessible because of classical computational limitations.

The algorithm enables simulations of a remarkably diverse range of quantum systems. For example, we have demonstrated the capability to accurately learn the ground states of local spin models in quantum magnetism and non-local electronic structure Hamiltonians, including cases with distorted molecular geometries dominated by strong multireference correlations. The versatility of this method opens up avenues for studying even more complex systems, including the following:

\begin{itemize}

\item \textbf{Exotic Topological Phases}: Our approach is well suited for exploring phase transitions in strongly correlated topological systems, such as symmetry-protected topological phases and magnetic topological orders \cite{chen2012symmetry, schuch2011classifying, sajjan2022magnetic}. Previous studies have demonstrated the capability of NQS, including restricted Boltzmann machines  and deep neural networks, to effectively model topological systems and investigate their phase transitions with high accuracy \cite{PhysRevB.99.155136, gao2017efficient}. These works highlight the utility of NQS in capturing intricate quantum correlations and topological order. Building on this foundation, our method aims to simulate systems with even more complex topological properties at larger scales in order to uncover emergent quantum phases of matter and better understand the interplay between order parameters and entanglement.

\item \textbf{Open Quantum Systems}: NQS have been used to model steady states of open quantum systems described by a Liouvillian, such as dissipative spin chains and cavity-QED systems \cite{luo2022autoregressive, PhysRevB.99.214306}. Our algorithm can enhance these studies by reducing sampling errors and enabling exploration of non-Markovian effects, such as memory-dependent dissipation in driven-dissipative Bose--Hubbard models \cite{PhysRevLett.125.115301}. In the future, we could explore the accurate modeling of memory effects in open quantum systems and scaling to larger networks of interacting modes.

\item \textbf{Finite-Temperature Physics}: Neural quantum states  have proven effective in representing density matrices, broadening their application from dissipative quantum systems to finite-temperature states of quantum many-body systems. This process typically involves representing the density matrix of a thermal state using NQS itself or sampling an ensemble of pure states according to the correct thermal distribution, enabling calculations of thermal averages by measuring observables across the ensemble \cite{PRXQuantum.2.010317, PhysRevLett.123.090402, powers2023exploring, motta2020determining}. NQS has been successfully applied to study finite-temperature properties of systems such as cold bosonic atoms in optical lattices \cite{PhysRevResearch.2.013284}, where intricate thermal behaviors emerge. Additionally, thermal effects, such as magnetic-field-driven phase transitions in high-temperature superconducting materials such as cuprates \cite{kemper2016thermodynamic}, represent compelling avenues for exploration. Our method can accelerate the exploration of critical thermal phase transitions, enabling accurate resolution of quantum criticality and thermal fluctuations in strongly correlated materials.

\item \textbf{Quantum Dynamics}: Our algorithm can facilitate the study of dynamical evolutions of steady states generated by a Hamiltonian or Liouvillian and can potentially simulate quench dynamics in out-of-equilibrium systems \cite{heyl2013dynamical, kloss2020studying, mendoza2022dynamical, budich2016dynamical}. Variational Monte Carlo approaches have previously been employed to simulate non-equilibrium steady states of open quantum systems \cite{hryniuk2024tensor}, offering a framework for investigating systems beyond equilibrium. Our method aims to address the current challenges in this domain, including  characterizing prethermalization phenomena \cite{mori2018thermalization,cazalilla2012thermalization} and capturing long-time dynamics in strongly interacting systems.

\end{itemize}

However, challenges remain in implementing such methods on NISQ devices:
\begin{itemize}
    \item \textbf{Hardware Limitations:} While the proposed algorithm minimizes circuit complexity, noise, and decoherence in quantum devices, it still imposes constraints on scalability and accuracy.
    \item \textbf{Measurement Overheads:} While our algorithm improves the accuracy of sampling from the desired distribution \( P(v) \) using polynomial resources, it does not address a fundamental limitation of sampling-based methods: the sample complexity of estimating expectation values depends on the alignment between the operator and the state. If \( P(v) \) is broadly distributed or the operator is supported on rare configurations, achieving an error \(\epsilon\) may still require an exponentially large number of samples.

\end{itemize}

Addressing these roadblocks can unlock the full potential of NQS training, enabling broader applications in quantum chemistry and condensed matter physics.

\subsection{Information Scrambling and Quantum Machine Learning}
The exploration of imaginary components of out-of-time-order correlators (OTOCs) \cite{PhysRevResearch.5.013146} to study information scrambling within quantum machine learning models offers profound insights into the learning dynamics and emulation capabilities of these systems. By connecting the imaginary OTOC components to quantum mutual information, this research highlights a novel avenue for quantifying the information scrambling capacity of quantum neural networks.

Opportunities identified in this work include the following:

\begin{itemize}
    \item \textbf{Enhanced Model Interpretability:}
    The construction of a convex space to track bounds on information scrambling and correlation dynamics offers a powerful tool for visualizing and understanding the training landscape of quantum neural networks. This goes beyond treating QML models as black boxes, fostering the development of theoretically sound and effective ansätze.

    \item \textbf{Phase Distinction in Physical Systems:} The ability of latent units in the network to mirror spin correlations and identify phase boundaries suggests potential applications in quantum simulation and materials discovery.
    \item \textbf{Scalability to Larger Systems:} The observed saturation of bounds in large physical systems underscores the robustness of the approach, paving the way for its application to industrial-scale problems.
\end{itemize}

Several roadblocks also need to be addressed:
\begin{itemize}
    \item \textbf{Complexity of Analytical Bounds:} While the derived bounds enhance theoretical understanding, translating these into actionable training metrics for complex quantum architectures remains a challenge.
    \item \textbf{Optimization Strategies:} Designing training protocols that maximize scrambling while ensuring convergence to desired solutions requires further investigation.
    \item \textbf{Phase Description:} The current method focuses on neural networks with real parameters, which cannot fully capture the phase information of quantum states. Extending these methods to complex-valued networks represents a critical next step.
\end{itemize}

Current findings open the door to new algorithms that exploit information scrambling as a feature, bridging the gap between physics-inspired measures and practical quantum model design.

\subsection{Tensor Networks and Classical Verification}
While our primary focus is to develop scalable quantum algorithms to simulate large, complex many-body systems and study their properties, as system size increases, verifying the accuracy of quantum simulation results becomes crucial. Therefore, in our group, we partake in complementary research on classical simulation using techniques such as tensor networks on classical high-performance computing platforms, which offer a robust verification method and benchmarking.

Tensor networks, particularly matrix product state  and projected entangled pair states, are highly tunable via bond dimension and can asymptotically approach exact results for many-body systems. (Methods exist to simulate dynamics, finite temperature, and open quantum systems, too.) This capability provides a benchmark for evaluating quantum simulation outcomes, ensuring the reliability of quantum algorithms in the large-scale limit.

\subsection{Error Mitigation and Error Correction}
Working with current quantum computers presents significant challenges due to the prevalence of noise and errors in NISQ devices. Cutting-edge error mitigation techniques should be developed and employed to alleviate these issues. These methods are essential for achieving meaningful results on NISQ platforms.

For certain applications, error mitigation alone is insufficient, necessitating the use of fault-tolerant quantum computers. Ongoing research in error correction aims to bridge this gap by exploring innovative techniques to improve the fidelity and reliability of quantum computations. Addressing these challenges will contribute to the foundational development of quantum algorithms, ensuring their scalability, accuracy, and practical applicability across diverse domains.

\section{Numerical Optimization}
\subsection*{Jeffrey Larson, Argonne National Laboratory, Lemont, IL, jmlarson@anl.gov}

The development and deployment of quantum algorithms face significant
roadblocks/impediments due to the lack of high-quality, specialized numerical optimization
methods and models. This challenge is especially evident in areas critical to
quantum algorithm performance, including parameter tuning, error correction, and
quantum circuit synthesis. We outline key roadblocks in these areas,
emphasizing their impact and the need for improvements.

\subsection{Stochastic Optimization for Parameter Identification}

Many quantum algorithms (QAOA, VQE, and adaptive variants such as ADAPT-VQE),
rely on parameterized quantum circuits whose efficacy depends
on high-quality parameter selection \cite{Cerezo_2021}. ADAPT-VQE, in particular, dynamically
constructs the ansatz during optimization, which introduces unique challenges
in identifying effective parameters for both ansatz growth and energy
minimization \cite{Grimsley_2019}. Numerical optimization methods are further complicated by noisy quantum
measurements and high-dimensional, nonconvex optimization landscapes. Current
stochastic optimization techniques lack the robustness and efficiency needed to
tackle these challenges, especially when dealing with limited quantum
resources and the relative expense of each objective evaluation \cite{Shaydulin_2019}. Developing
optimization methods that can efficiently navigate noisy landscapes while
adapting to evolving circuit structures is critical to improving algorithmic
performance and scalability.

\subsection{Discrete Optimization for Error Correction}

Error correction remains a cornerstone for achieving fault-tolerant quantum
computing. Optimizing the mapping of error-correcting codes, such as quantum
low-density parity-check codes, to hardware presents significant
challenges to discrete optimization methods \cite{Bravyi_2024}. Existing tools fail to scale to the complexity of modern
quantum hardware and often do not exploit the specific structure of these
problems \cite{Poole_2024}. Advancing discrete optimization techniques that incorporate
domain-specific knowledge and are tailored for quantum systems is necessary to
make error correction more practical and efficient.

\subsection{Optimization for Quantum Circuit Synthesis}

The efficacy of quantum algorithms depends heavily on the quality of their
circuit representations, particularly the number of two-qubit gates (e.g., CNOT
gates) and overall circuit depth \cite{Giles_2013}. Reducing circuit depth is crucial for
mitigating noise and enhancing performance, particularly when comparing quantum
algorithms with their classical counterparts. However, current circuit synthesis
methods often fall short of producing optimal circuits for large-scale or
hardware-specific implementations \cite{Liu_2024}. New approaches that combine numerical
optimization with emerging techniques in hybrid classical-quantum workflows are
essential. These methods should target efficient gate decomposition and
hardware-aware synthesis to optimize both execution time and fidelity.

\subsection{Optimization Challenges in Simulations and Machine Learning for Quantum Systems}

Emerging applications of quantum algorithms, such as simulations of quantum
systems and machine learning, introduce new optimization roadblocks. For
example, Hamiltonian learning and state preparation often require solving
highly nonconvex problems with limited measurements and substantial noise.
Talks presented at the Joint Algorithms Workshop (i.e., those on quantum
control, hybrid quantum-classical frameworks, and algorithmic benchmarking)
underscore the need for optimization techniques that handle these multifaceted
challenges. Developing methods that can exploit quantum-specific symmetries,
adapt to hybrid architectures, and integrate seamlessly with machine learning
frameworks will be vital for enabling these advanced applications.
Additionally, improving benchmarking methodologies to assess algorithmic
performance quantitatively is necessary to guide optimization strategies and
ensure competitiveness with classical approaches.

\subsection{Needed: A Coordinated Effort in Advanced Optimization}

The lack of advanced numerical optimization methods hampers the
progress of quantum algorithm research and its translation to real-world
applications \cite{hoefler2023disentangling}. Addressing these roadblocks requires a coordinated effort to
develop optimization methods that exploit the unique characteristics of quantum
problems, are robust to noise, and scale effectively with problem size. By
overcoming these challenges, we can unlock the full potential of quantum
algorithms and position them as competitive alternatives to classical methods
in diverse domains.

\subsection*{Acknowledgments}
This material is based upon work supported by Q-NEXT, one of the U.S.
Department of Energy Office of Science National Quantum Information Science
Research Centers.

\section{Early Fault-Tolerant Quantum Computing and Quantum Sensing}

\subsection*{Andrew Sornborger, Los Alamos National Laboratory, Los Alamos, NM, sornborg@lanl.gov}

\subsection{Recent Challenges in Quantum Machine Learning}

Two recent developments  are changing my research foci in quantum computation. The first is the recent review linking the absence of barren plateaus in quantum machine learning and variational quantum algorithms with classical simulability \cite{larocca2025barren} (and the many citations therein). Barren plateaus are problematic for quantum machine learning, since they lead to an optimizer's inability to determine the direction to a cost minimum and hence the inability to effectively train parameters. Although not yet generally proven (but a number of instances have been), this link makes it difficult to envision scalable quantum machine learning algorithms of value in the current era before we have fault-tolerant hardware to play with. I say ``not yet," as opposed to ``not ever," because similar vanishing gradients and the curse of dimensionality exist for classical machine learning, but amazing progress has been made there. In the absence of fault tolerance, however, it is hard to envision an ability to play with heuristics that might be of value for quantum machine learning.

\subsection{Opportunities Resulting from What We Have Learned from Quantum Machine Learning}

Having decided that it is not yet time to pursue quantum machine learning, what other directions can we pursue with non-fault-tolerant hardware? My choice has been quantum sensing. Quantum sensing is an area where few-spin (qubit) systems have been little studied (relative to materials-sized systems or single-spin systems). Quantum sensing is also an area where the sensor response is effectively a cost function measuring the departure of a (small) quantum system from its background state due to an external signal. These features overlap considerably with quantum machine learning. But, instead of trying to start away from a minimum, then variationally modify parameters to find the minimum, we start at the minimum (no signal) and allow the system to drift away from the minimum (due to a signal). Thus, we are in some sense using the reverse procedure to quantum machine learning to detect a signal of interest with a quantum sensor. But the range of analytical techniques that we have for understanding variational algorithms should be valuable for quantum sensing. For coherent many-body quantum sensors, many  cost functions arising from quantum machine learning could be useful to study.

\subsection{Recent Developments in Fault Tolerance}

The second recent development that is changing my research focus is the many indications over the past two or three years that quantum hardware is at a tipping point, where qubit fidelities have reached the threshold needed to achieve valuable fidelity improvements by implementing quantum error correction to realize logical qubits \cite{ryan2021realization,google2023suppressing,sivak2023real,ryan2024high,bluvstein2024logical,gupta2024encoding}.

At LANL, we used to joke that one of the first quantum algorithms to give quantum advantage could be quantum error correction itself. This may, indeed, now be the case. Only a handful of quantum algorithm implementations can claim to surpass classical capabilities. And there is a continuing battle between classical and quantum algorithms making the point of distinction difficult to isolate because it changes so quickly. From this point of view,  a really good research target moving forward would seem to be to optimize quantum error correction for the various hardware architectures making use of the broad range of controllable quantum systems available. In the area of fault tolerance, many target problems exist that are both hardware-dependent and more abstract hardware-agnostic issues in optimization and machine learning.

\subsection{Opportunities for Making Fault-Tolerant Systems More Efficient}

A glance at the arXiv repository  clearly indicates that much of the field is well aware of the developments in fault-tolerant hardware already, so this statement will not be surprising to anyone. But it is important to consider what research directions will be most fruitful for different participants in the community. Clearly, industry---particularly large, well-funded companies---have already developed comprehensive roadmaps for advancing fault tolerance for their hardware. Opportunities remain, however, for instance for DOE national laboratories to work in co-design teams with industry to optimize particular algorithms for mission purposes. Quantum simulation is of major relevance to  DOE and will likely be a core focus. Individual researchers at universities will have the opportunity to team up with larger teams or to focus on individual components of fault tolerance.

\subsection*{Acknowledgments}

This material is based upon work supported by the U.S. Department of Energy, Office of Science, National Quantum Information Science Research Centers, Quantum Science Center (QSC).

\section{Establishment of Quantum Advantage}
\subsection*{Peter Love, Department of Physics and Astronomy, Tufts University (Quantum Systems Accelerator), Medford,MA, peter.love@tufts.edu}
The existence of NISQ devices that are out of reach of direct classical simulation poses the question of quantum advantage: When will a  calculation be performed on a quantum computer that cannot be performed on a classical computer? This simple question is unfortunately vaguer than it appears.  In this short article we argue that a roadblock to establishing claims of advantage is the lack of a clear protocol proposing, claiming, and challenging claims of quantum advantage. We make some suggestions as to the form such a protocol might take.

Quantum computers capable of performing simple algorithms on a small scale have existed for over a decade. For example, many small calculations in the area of quantum simulation represent the first application of quantum computation to a particular area or the first use of a new method~\cite{lanyon2010towards,peruzzo2014variational,wang2015quantum,kandala2017hardware,o2016scalable,hempel2018quantum,dumitrescu2018cloud,google2020hartree,kreshchuk2021light}. Demonstrations on a few tens of qubits may be simulated directly by modest classical computers. The question of the utility of such demonstrations may be answered by Faraday's famous remark ``Of what use is a newborn baby?''~\footnote{Exactly when Faraday said this and to whom is not known, although it was in response to a question about magnetic induction, and so it must have been after 1831.}

These early experiments heralded the NISQ era~\cite{preskill2018quantum}, which arguably began in earnest with the first demonstration of quantum supremacy in 2019~\cite{arute2019quantum}. This quantum supremacy demonstration established with a high degree of confidence that the Sycamore device on which the demonstration was performed was out of reach of direct classical simulation. Unfortunately, the problem solved in supremacy experiments is not useful for applications other than demonstrating the power of a particular quantum device.

\subsection{Supremacy vs. Quantum Advantage}
To continue discussion of the performance of current and future quantum computers, we need to clarify some terminology. In this article we forgo any qualms about the term ``supremacy'' and use it to refer to any clear, but practically useless, demonstration of superiority of quantum computers over classical. We use the term ``quantum advantage'' for demonstrations of advantage that apply to useful applications. This notion of usefulness itself can be vague, and so we use the term ``quantum advantage'' to refer to any calculation for which use of a quantum computer would be preferable to use of a classical computer. This preference may arise from any metric---time to solution, quality of solution, energy requirements, and so on.~\footnote{This criterion is also sometimes called quantum utility, but we consider advantage to be clearer because it better reflects the competition between classical and quantum approaches.}

Demonstration of quantum supremacy is neither necessary nor sufficient to establish advantage. However, the ideas and motivations behind supremacy experiments should remind us that we are not merely engaged in the business of building a better quantum  mousetrap. Quantum computing also shakes the foundations of computer science by challenging the extended Church--Turing thesis that all implementations of computation are equivalently powerful~\cite{turing1936computable,church1936unsolvable}. Quantum supremacy marks a radical, but not practical, break in the nature of computation. The demonstration of quantum advantage would show a qualitative difference in practice between classical and quantum computation, indicating that the development of theoretical computer science prior to quantum computing is merely a preamble, in the same way that ray optics was a preamble to the wave theory of light. In the remainder of this article we will argue that establishing claims of advantage should be treated with the same seriousness as claims of other new physical phenomena such as dark matter or gravitational waves.

\subsection{Algorithmic vs. Practical Advantage}
Quantum advantage must begin with an algorithmic advantage: current quantum technology is unlikely to evolve to have faster rates of elementary operations, and so advantage must arise from requiring fewer quantum operations than classical operations for the same task.  Algorithmic advantage for specific tasks is relatively clear-cut: if the growth of quantum resources with problem size is slower than the growth of classical resources for sufficiently large problem size, we will achieve a quantum advantage.  With access to a large-scale reliable quantum computer one could simply run one's favorite algorithm and observe the improved scaling. The prospects for doing so in practice have been reviewed in two recent papers~\cite{hoefler2023disentangling,babbush2021focus}. These works emphasize the distinction between algorithmic and practical advantage. One simple way of stating this distinction is that the overhead of realizing a problem instance on a quantum computer may push the advantage regime into instance sizes that are well beyond those of actual practical interest.

Prior to the availability of NISQ devices with many tens or hundreds of qubits, practical demonstrations of advantage were simply not possible. Direct simulation of the device itself at qubit numbers below around $50$ precludes any advantage demonstration, and scaling demonstrations based on extrapolation from $N\leq50$ cannot be convincing. With the demonstration of quantum supremacy~\cite{arute2019quantum}, we have evidence that NISQ devices have reached a scale at which direct classical simulation is at least overwhelmingly difficult. These devices have also enabled practical calculations at large scale---two recent examples in chemistry and lattice gauge theory reached well beyond the $50$ qubit scale~\cite{robledo2024chemistry,farrell2024scalable}. These large-scale practical NISQ demonstrations are encouraging. However, using a quantum computer to solve a problem  that is too large to be directly classically simulated is not sufficient to establish advantage (and~\cite{robledo2024chemistry,farrell2024scalable} do not claim advantage). Quantum calculations must outperform all classical methods, which can exploit problem structure.  This point has been sharply made for chemistry in~\cite{lee2023evaluating}.

Large-scale quantum computing calculations provide rich targets for classical computational science. Several large-scale demonstrations have been rapidly challenged by classical methods. Quantum supremacy experiments~\cite{boixo2018characterizing,arute2019quantum} have been challenged in various ways~\cite{aharonov2023polynomial,pednault2019leveraging,oh2023spoofing}. A NISQ calculation of the transverse Ising model in two dimensions~\cite{kim2023evidence} showed advantage over ``brute-force'' methods, but subsequent classical calculations taking more advantage of problem structure were able to reproduce these results~\cite{tindall2024efficient,beguvsic2024fast,liao2023simulation}. The continuous-time dynamics of two-dimensional Ising models was recently simulated using a quantum annealer~\cite{king2023quantum,king2025beyond}, and similarly challenged by classical simulation~\cite{tindall2025dynamics,mauron2025challenging,park2025simulating,king2025comment}.

\subsection{Criteria for Claim of Quantum Advantage}
These considerations illustrate why establishing a claim of quantum advantage is a difficult problem.  A claim of quantum advantage compares quantum and classical costs for actual computations on a single instance or set of instances of a particular problem. For such a claim to be made, the quantum costs, including all overheads of mapping the problem, must be  optimized. For the claim to stand up, the classical costs must also be optimized.  Achieving optimal realizations of both classical and quantum methods is extremely difficult technically for a single group. There is also an obvious problem of motivation for a mainly quantum computing group to work hard on classical techniques in order to disprove the utility of their devices. From these considerations one concludes that establishing quantum advantage is an inherently {\em adversarial} problem, that classical and quantum realizations should be performed by different groups.  Indeed, in the examples discussed above, this is what is happening as part of the natural process of science: claims are examined, challenged, and revised. In the process new classical algorithms have been discovered that move the advantage frontier forward. Advantage will be achieved when quantum calculations reach sufficient scale and when classical ingenuity has been exhausted.

\subsection{Targets for Quantum Advantage}
We propose that this virtuous cycle should be formalized by establishing protocols for claims of advantage, drawing lessons from similar problems in other fields. These are only suggestions, and we anticipate broader engagement of the community with this question will improve upon them. First, we consider how to improve communication about targets for quantum advantage. It is useful to draw upon techniques aimed at solving the reproducibility crisis~\cite{baker20161}---registered reports are particularly interesting~\cite{chambers2022past}. In the context of supremacy a registered report would publish a target calculation, as well as the quantum algorithm and implementation proposed to solve it, before performing the experiment. Full details would therefore be available to all competing groups, classical as well as quantum. Classical and quantum solutions could then be published and compared equitably. If rapid classical solution is in fact possible, the proposed advantage target could be revised to reflect this, before effort is expended to perform the quantum calculation. The collection of instances in these registered reports would eventually form a collection of hard instances for all classical algorithms. These instances would constitute a benchmark set similar to the ``bad chemistry'' examples in the NIST CCCDB~\cite{CCCBD}.

\subsection{Market Mechanisms}
Beyond using new means of communication about targets for quantum advantage, formal adversarial mechanisms could be created around these targets. One example is competitions, such as the SAT solver competitions with a long history~\cite{buro1992report,simon2005sat2002,le2003essentials,kautz2004walksat,kautz2004walksat,jarvisalo2012international,balyo2017sat,heule2019sat,kochemazov2020improving}. These competitions provide clear and open evaluations of existing and new algorithms, where the performance of each algorithm on a common set of instances is documented in published proceedings.

It is natural to turn to economics when attempting to establish productive mechanisms in adversarial settings.
The interplay between economics and computational science has been a rich area of activity for some time. Using ideas from computational science to solve problems in markets has led to optimization of returns from auctions, to optimization of trading of electricity resources, and to improved portfolio optimization. Use of economic ideas in computational science has led to market-based management of cloud and grid computing services. Portfolio theory applied to algorithms has been used to produce solutions to NP-complete problems more quickly ~\cite{huberman1997economics}. Computational complexity has also been used to show that markets themselves are not efficient unless P=NP~\cite{efficientNP}.

Markets serve as a basic coordinating mechanism between agents whose cooperation may not be relied upon. As such, they can play a useful role in large scientific endeavors with many actors who may or may not have sufficient incentives to cooperate. One example of such a set of actors is computational scientists deploying algorithms in an application domain. Algorithm evaluation is difficult, because typically practitioners choose the problem instances to study, which can lead to selection bias in the performance analysis of the algorithms. Such performance evaluation is best done adversarially, but this requires optimal use of an algorithm by an actor who is trying to extract worst-case performance. It is of course always easy to obtain poor results by inefficient implementation or operation of an algorithm.

A market mechanism is needed to stimulate algorithmic research and to identify best-in-class algorithms in an open but adversarial context ~\cite{problemmarkets}. The work of~\cite{problemmarkets} proposed a securities market, in which one can buy a security that pays $\$1$ if a solution to a particular instance of an NP-complete problem is found by a particular date. Clearly, if the instance is known to be easy to solve, the price of the security should be $\$1$---anyone can buy it, solve the instance, and receive $\$1$. The lower the price of the security, the harder the market believes the instance to be, and the greater the reward for solving it. To reap the reward, one simply buys large numbers of cheap securities for the instance and then solves the corresponding instances. Such markets require a trusted source of problem instances, because it is possible to construct instances that have a ``planted'' solution that are nonetheless challenging to solve a priori. This idea was explored for satisfiability, satisfiability counting, and Bayesian inference in~\cite{problemmarkets}. One could imagine establishing a market in the underlying problems of interest. These could be quantum chemistry, quantum computing applied to high energy physics problems, and combinatorial optimization problems. The ``price'' of an instance will be set by the best possible solution produced by classical or quantum means.

The underlying market in problem instances would therefore provide a valuable coordination mechanism by itself. Building a derivatives market on top of the underlying market is straightforward by the method outlined in~\cite{problemmarkets} . A derivatives market for simulation problems in physical science such as quantum chemistry or high energy physics poses some additional challenges related to validation of the solutions. One could initially focus on problems that are variational, such as ground state energy estimation where the best known solution serves as an upper bound to future possible solutions. The conversion of ground state estimation problems to decision problems is well understood in quantum information through the formulation of local Hamiltonian as a QMA-complete problem. A derivative on such instances could therefore constitute a bet that the energy is below a certain threshold, and the price of such securities as a function of the threshold would represent the belief of the market concerning the true ground state energy.

The market would function as follows. Just as in publicly funded supercomputer resources, users could apply for access and receive a certain amount of financing in the form of tokens. Users can earn tokens by uploading instances to the market or by solving instances or by engaging in trading on the derivatives market. While no actual money would be exchanged, the market could be used as a mechanism to validate and identify achievements on the algorithmic front, either quantum or classical, and promote these achievements.

\subsection{More Productive Communication Needed}
In this article we have argued that reliably establishing quantum advantage is challenging both technically and sociologically. We have suggested several mechanisms that could facilitate communication about  targets and accomplishments and hopefully make the interaction between advances in classical and quantum algorithms more productive.

\bibliographystyle{unsrtnat}
\bibliography{Kapit,Chong,Crane,Kais,Larson,Sornborger,love}

\begin{thebibliography}{172}
\providecommand{\natexlab}[1]{#1}
\providecommand{\url}[1]{\texttt{#1}}
\expandafter\ifx\csname urlstyle\endcsname\relax
  \providecommand{\doi}[1]{doi: #1}\else
  \providecommand{\doi}{doi: \begingroup \urlstyle{rm}\Url}\fi

\bibitem[Korte et~al.(2011)Korte, Vygen, Korte, and
  Vygen]{korte2011combinatorial}
Bernhard~H Korte, Jens Vygen, B~Korte, and J~Vygen.
\newblock \emph{Combinatorial Optimization}, volume~1.
\newblock Springer, 2011.

\bibitem[M{\'e}zard et~al.(2002)M{\'e}zard, Parisi, and
  Zecchina]{mezard2002analytic}
Marc M{\'e}zard, Giorgio Parisi, and Riccardo Zecchina.
\newblock Analytic and algorithmic solution of random satisfiability problems.
\newblock \emph{Science}, 297\penalty0 (5582):\penalty0 812--815, 2002.

\bibitem[M{\'e}zard et~al.(2005)M{\'e}zard, Mora, and
  Zecchina]{mezard2005clustering}
Marc M{\'e}zard, Thierry Mora, and Riccardo Zecchina.
\newblock Clustering of solutions in the random satisfiability problem.
\newblock \emph{Physical Review Letters}, 94\penalty0 (19):\penalty0 197205,
  2005.

\bibitem[Hartmann and Weigt(2006)]{hartmann2006phase}
Alexander~K Hartmann and Martin Weigt.
\newblock \emph{Phase Transitions in Combinatorial Optimization Problems:
  Basics, Algorithms and Statistical Mechanics}.
\newblock John Wiley \& Sons, 2006.

\bibitem[Krzakala and Kurchan(2007)]{krzakala2007landscape}
Florent Krzakala and Jorge Kurchan.
\newblock Landscape analysis of constraint satisfaction problems.
\newblock \emph{Physical Review E}, 76\penalty0 (2):\penalty0 021122, 2007.

\bibitem[Altarelli et~al.(2008)Altarelli, Monasson, and
  Zamponi]{altarelli2008relationship}
Fabrizio Altarelli, R{\'e}mi Monasson, and Francesco Zamponi.
\newblock Relationship between clustering and algorithmic phase transitions in
  the random k-xorsat model and its {NP}-complete extensions.
\newblock In \emph{Journal of Physics: Conference Series}, volume~95, page
  012013. IOP Publishing, 2008.

\bibitem[Gamarnik(2021)]{gamarnik2021overlap}
David Gamarnik.
\newblock The overlap gap property: A topological barrier to optimizing over
  random structures.
\newblock \emph{Proceedings of the National Academy of Sciences}, 118\penalty0
  (41):\penalty0 e2108492118, 2021.

\bibitem[Kapit et~al.(2024)]{kapit2024approximability}
Eliot Kapit et~al.
\newblock On the approximability of random-hypergraph max-3-xorsat problems
  with quantum algorithms.
\newblock \emph{arXiv preprint arXiv:2312.06104}, 2024.

\bibitem[Jordan et~al.(2024)Jordan, Shutty, Wootters, Zalcman, Schmidhuber,
  King, Isakov, and Babbush]{jordan2024optimization}
Stephen~P Jordan, Noah Shutty, Mary Wootters, Adam Zalcman, Alexander
  Schmidhuber, Robbie King, Sergei~V Isakov, and Ryan Babbush.
\newblock Optimization by decoded quantum interferometry.
\newblock \emph{arXiv preprint arXiv:2408.08292}, 2024.

\bibitem[Hastings(2020)]{hastings2020classical}
Matthew~B Hastings.
\newblock Classical and quantum algorithms for tensor principal component
  analysis.
\newblock \emph{Quantum}, 4:\penalty0 237, 2020.

\bibitem[Schmidhuber et~al.(2024)Schmidhuber, O'Donnell, Kothari, and
  Babbush]{schmidhuber2024quartic}
Alexander Schmidhuber, Ryan O'Donnell, Robin Kothari, and Ryan Babbush.
\newblock Quartic quantum speedups for planted inference.
\newblock \emph{arXiv preprint arXiv:2406.19378}, 2024.

\bibitem[Barton et~al.(2024)Barton, Sagal, Feeney, Grattan, Patnaik, Oganesyan,
  Carr, and Kapit]{barton2024iterative}
Brandon Barton, Jacob Sagal, Sean Feeney, George Grattan, Pratik Patnaik, Vadim
  Oganesyan, Lincoln~D Carr, and Eliot Kapit.
\newblock Iterative quantum optimization of spin glass problems with rapidly
  oscillating transverse fields.
\newblock \emph{arXiv preprint arXiv:2408.06571}, 2024.

\bibitem[Zalka(1999)]{zalka1999}
Christof Zalka.
\newblock {Grover}'s quantum searching algorithm is optimal.
\newblock \emph{Physical Review A}, 60\penalty0 (4):\penalty0 2746, 1999.

\bibitem[Aaronson and Ambainis(2009)]{aaronson2009need}
Scott Aaronson and Andris Ambainis.
\newblock The need for structure in quantum speedups.
\newblock \emph{arXiv preprint arXiv:0911.0996}, 2009.

\bibitem[Montanaro and Zhou(2024)]{montanaro2024quantum}
Ashley Montanaro and Leo Zhou.
\newblock Quantum speedups in solving near-symmetric optimization problems by
  low-depth {QAOA}.
\newblock \emph{arXiv preprint arXiv:2411.04979}, 2024.

\bibitem[Arora and Barak(2009)]{arora2009computational}
Sanjeev Arora and Boaz Barak.
\newblock \emph{Computational Complexity: A Modern Approach}.
\newblock Cambridge University Press, 2009.

\bibitem[Bellitti et~al.(2021)Bellitti, Ricci-Tersenghi, and
  Scardicchio]{bellitti2021entropic}
Matteo Bellitti, Federico Ricci-Tersenghi, and Antonello Scardicchio.
\newblock Entropic barriers as a reason for hardness in both classical and
  quantum algorithms, 2021.

\bibitem[Farhi et~al.(2014)Farhi, Goldstone, and Gutmann]{farhi2014quantum}
Edward Farhi, Jeffrey Goldstone, and Sam Gutmann.
\newblock A quantum approximate optimization algorithm.
\newblock \emph{arXiv preprint arXiv:1411.4028}, 2014.

\bibitem[Albash and Lidar(2017)]{albashlidar2017}
Tameem Albash and Daniel~A. Lidar.
\newblock Adiabatic quantum computing.
\newblock \emph{arXiv:1611.04471}, 2017.

\bibitem[Babbush et~al.(2021)Babbush, McClean, Newman, Gidney, Boixo, and
  Neven]{babbush2021focus}
Ryan Babbush, Jarrod~R McClean, Michael Newman, Craig Gidney, Sergio Boixo, and
  Hartmut Neven.
\newblock Focus beyond quadratic speedups for error-corrected quantum
  advantage.
\newblock \emph{PRX Quantum}, 2\penalty0 (1):\penalty0 010103, 2021.

\bibitem[Ramesh et~al.(2024)Ramesh, Tomesh, Riesenfeld, Chong, and
  Pearson]{ramesh2024quantum}
Siddhi Ramesh, Teague Tomesh, Samantha~J Riesenfeld, Frederic~T Chong, and
  Alexander~T Pearson.
\newblock Quantum computing for oncology.
\newblock \emph{Nature Cancer}, pages 1--6, 2024.

\bibitem[Echle et~al.(2021)Echle, Rindtorff, Brinker, Luedde, Pearson, and
  Kather]{echle2021deep}
Amelie Echle, Niklas~Timon Rindtorff, Titus~Josef Brinker, Tom Luedde,
  Alexander~Thomas Pearson, and Jakob~Nikolas Kather.
\newblock Deep learning in cancer pathology: a new generation of clinical
  biomarkers.
\newblock \emph{British Journal of Cancer}, 124\penalty0 (4):\penalty0
  686--696, 2021.

\bibitem[Howard et~al.(2021)Howard, Dolezal, Kochanny, Schulte, Chen, Heij,
  Huo, Nanda, Olopade, Kather, et~al.]{howard2021impact}
Frederick~M Howard, James Dolezal, Sara Kochanny, Jefree Schulte, Heather Chen,
  Lara Heij, Dezheng Huo, Rita Nanda, Olufunmilayo~I Olopade, Jakob~N Kather,
  et~al.
\newblock The impact of site-specific digital histology signatures on deep
  learning model accuracy and bias.
\newblock \emph{Nature Communications}, 12\penalty0 (1):\penalty0 4423, 2021.

\bibitem[Fan et~al.(2014)Fan, Han, and Liu]{fan_2014}
Jianqing Fan, Fang Han, and Han Liu.
\newblock {Challenges of Big Data analysis}.
\newblock \emph{National Science Review}, 1\penalty0 (2):\penalty0 293--314, 02
  2014.
\newblock ISSN 2095-5138.
\newblock \doi{10.1093/nsr/nwt032}.
\newblock URL \url{https://doi.org/10.1093/nsr/nwt032}.

\bibitem[Hrinivich et~al.(2023)Hrinivich, Wang, and Wang]{hrinivich_2023}
William~Thomas Hrinivich, Tonghe Wang, and Chunhao Wang.
\newblock Editorial: Interpretable and explainable machine learning models in
  oncology.
\newblock \emph{Frontiers in Oncology}, 13, 2023.
\newblock ISSN 2234-943X.
\newblock \doi{10.3389/fonc.2023.1184428}.
\newblock URL
  \url{https://www.frontiersin.org/journals/oncology/articles/10.3389/fonc.2023.1184428}.

\bibitem[Watkinson et~al.(2008)Watkinson, Wang, Zheng, and
  Anastassiou]{watkinson2008identification}
John Watkinson, Xiaodong Wang, Tian Zheng, and Dimitris Anastassiou.
\newblock Identification of gene interactions associated with disease from gene
  expression data using synergy networks.
\newblock \emph{BMC Systems Biology}, 2:\penalty0 1--16, 2008.

\bibitem[Watkinson et~al.(2009)Watkinson, Liang, Wang, Zheng, and
  Anastassiou]{watkinson2009inference}
John Watkinson, Kuo-ching Liang, Xiadong Wang, Tian Zheng, and Dimitris
  Anastassiou.
\newblock Inference of regulatory gene interactions from expression data using
  three-way mutual information.
\newblock \emph{Annals of the New York Academy of Sciences}, 1158\penalty0
  (1):\penalty0 302--313, 2009.

\bibitem[Boehm et~al.(2022)Boehm, Khosravi, Vanguri, Gao, and
  Shah]{boehm2022harnessing}
Kevin~M Boehm, Pegah Khosravi, Rami Vanguri, Jianjiong Gao, and Sohrab~P Shah.
\newblock Harnessing multimodal data integration to advance precision oncology.
\newblock \emph{Nature Reviews Cancer}, 22\penalty0 (2):\penalty0 114--126,
  2022.

\bibitem[Bravyi et~al.(2020)Bravyi, Kliesch, Koenig, and
  Tang]{bravyi2020obstacles}
Sergey Bravyi, Alexander Kliesch, Robert Koenig, and Eugene Tang.
\newblock Obstacles to variational quantum optimization from symmetry
  protection.
\newblock \emph{Physical Review Letters}, 125\penalty0 (26):\penalty0 260505,
  2020.

\bibitem[Grover(1996)]{grover1996fast}
Lov~K Grover.
\newblock A fast quantum mechanical algorithm for database search.
\newblock In \emph{Proceedings of the Twenty-Eighth Annual ACM Symposium on
  Theory of Computing}, pages 212--219, 1996.

\bibitem[Yamakawa and Zhandry(2024)]{yamakawa2024verifiable}
Takashi Yamakawa and Mark Zhandry.
\newblock Verifiable quantum advantage without structure.
\newblock \emph{Journal of the ACM}, 71\penalty0 (3):\penalty0 1--50, 2024.

\bibitem[Karp(1975)]{karp1975computational}
Richard~M Karp.
\newblock On the computational complexity of combinatorial problems.
\newblock \emph{Networks}, 5\penalty0 (1):\penalty0 45--68, 1975.

\bibitem[Gomes et~al.(2008)Gomes, Kautz, Sabharwal, and
  Selman]{gomes2008satisfiability}
Carla~P Gomes, Henry Kautz, Ashish Sabharwal, and Bart Selman.
\newblock Satisfiability solvers.
\newblock \emph{Foundations of Artificial Intelligence}, 3:\penalty0 89--134,
  2008.

\bibitem[Tomesh et~al.(2022)Tomesh, Gokhale, Omole, Ravi, Smith, Viszlai, Wu,
  Hardavellas, Martonosi, and Chong]{tomesh2022supermarq}
Teague Tomesh, Pranav Gokhale, Victory Omole, Gokul~Subramanian Ravi, Kaitlin~N
  Smith, Joshua Viszlai, Xin-Chuan Wu, Nikos Hardavellas, Margaret~R Martonosi,
  and Frederic~T Chong.
\newblock Supermarq: A scalable quantum benchmark suite.
\newblock In \emph{2022 IEEE International Symposium on High-Performance
  Computer Architecture (HPCA)}, pages 587--603. IEEE, 2022.

\bibitem[Lubinski et~al.(2023)Lubinski, Johri, Varosy, Coleman, Zhao, Necaise,
  Baldwin, Mayer, and Proctor]{lubinski2023application}
Thomas Lubinski, Sonika Johri, Paul Varosy, Jeremiah Coleman, Luning Zhao,
  Jason Necaise, Charles~H Baldwin, Karl Mayer, and Timothy Proctor.
\newblock Application-oriented performance benchmarks for quantum computing.
\newblock \emph{IEEE Transactions on Quantum Engineering}, 4:\penalty0 1--32,
  2023.

\bibitem[Lubinski et~al.(2024)Lubinski, Coffrin, McGeoch, Sathe, Apanavicius,
  Bernal~Neira, Consortium, et~al.]{lubinski2024optimization}
Thomas Lubinski, Carleton Coffrin, Catherine McGeoch, Pratik Sathe, Joshua
  Apanavicius, David Bernal~Neira, Quantum Economic~Development Consortium,
  et~al.
\newblock Optimization applications as quantum performance benchmarks.
\newblock \emph{ACM Transactions on Quantum Computing}, 5\penalty0
  (3):\penalty0 1--44, 2024.

\bibitem[Yasuoka(2021)]{yasuoka2021computational}
Hirotoshi Yasuoka.
\newblock Computational complexity of quadratic unconstrained binary
  optimization.
\newblock \emph{arXiv preprint arXiv:2109.10048}, 2021.

\bibitem[Shaydulin et~al.(2024)Shaydulin, Li, Chakrabarti, DeCross, Herman,
  Kumar, Larson, Lykov, Minssen, Sun, et~al.]{shaydulin2024evidence}
Ruslan Shaydulin, Changhao Li, Shouvanik Chakrabarti, Matthew DeCross, Dylan
  Herman, Niraj Kumar, Jeffrey Larson, Danylo Lykov, Pierre Minssen, Yue Sun,
  et~al.
\newblock Evidence of scaling advantage for the quantum approximate
  optimization algorithm on a classically intractable problem.
\newblock \emph{Science Advances}, 10\penalty0 (22):\penalty0 eadm6761, 2024.

\bibitem[Abbas et~al.(2024)Abbas, Ambainis, Augustino, B{\"a}rtschi, Buhrman,
  Coffrin, Cortiana, Dunjko, Egger, Elmegreen, et~al.]{abbas2024challenges}
Amira Abbas, Andris Ambainis, Brandon Augustino, Andreas B{\"a}rtschi, Harry
  Buhrman, Carleton Coffrin, Giorgio Cortiana, Vedran Dunjko, Daniel~J Egger,
  Bruce~G Elmegreen, et~al.
\newblock Challenges and opportunities in quantum optimization.
\newblock \emph{Nature Reviews Physics}, pages 1--18, 2024.

\bibitem[Holevo(1973)]{holevo1973bounds}
Alexander~Semenovich Holevo.
\newblock Bounds for the quantity of information transmitted by a quantum
  communication channel.
\newblock \emph{Problemy Peredachi Informatsii}, 9\penalty0 (3):\penalty0
  3--11, 1973.

\bibitem[Hieromnimon et~al.(2023)Hieromnimon, Dolezal, Doytcheva, Howard,
  Kochanny, Zhang, Grossman, Tanager, Wang, Kather,
  et~al.]{hieromnimon2023latent}
Hanna~M Hieromnimon, James Dolezal, Kristina Doytcheva, Frederick~M Howard,
  Sara Kochanny, Zhenyu Zhang, Robert~L Grossman, Kevin Tanager, Cindy Wang,
  Jakob~Nikolas Kather, et~al.
\newblock Latent transcriptional programs reveal histology-encoded tumor
  features spanning tissue origins.
\newblock \emph{bioRxiv}, pages 2023--03, 2023.

\bibitem[Hoefler et~al.(2023)Hoefler, H{\"a}ner, and
  Troyer]{hoefler2023disentangling}
Torsten Hoefler, Thomas H{\"a}ner, and Matthias Troyer.
\newblock Disentangling hype from practicality: On realistically achieving
  quantum advantage.
\newblock \emph{Communications of the ACM}, 66\penalty0 (5):\penalty0 82--87,
  2023.

\bibitem[Arrasmith et~al.(2021)Arrasmith, Babbush, Benjamin, Endo, Fujii,
  McClean, Mitarai, Yuan, Cincio, et~al.]{cerezo2021variational}
Andrew Arrasmith, Ryan Babbush, Simon~C Benjamin, Suguru Endo, Keisuke Fujii,
  Jarrod~R McClean, Kosuke Mitarai, Xiao Yuan, Lukasz Cincio, et~al.
\newblock Variational quantum algorithms.
\newblock \emph{Nature Reviews Physics}, 3\penalty0 (9):\penalty0 625--644,
  2021.

\bibitem[Galda et~al.(2021)Galda, Liu, Lykov, Alexeev, and
  Safro]{galda_transferability_2021}
Alexey Galda, Xiaoyuan Liu, Danylo Lykov, Yuri Alexeev, and Ilya Safro.
\newblock Transferability of optimal {QAOA} parameters between random graphs.
\newblock \emph{arXiv:2106.07531 [quant-ph]}, June 2021.
\newblock URL \url{http://arxiv.org/abs/2106.07531}.
\newblock arXiv: 2106.07531.

\bibitem[Shaydulin et~al.(2023)Shaydulin, Lotshaw, Larson, Ostrowski, and
  Humble]{shaydulin_parameter_2023}
Ruslan Shaydulin, Phillip~C. Lotshaw, Jeffrey Larson, James Ostrowski, and
  Travis~S. Humble.
\newblock Parameter transfer for quantum approximate optimization of weighted
  {MaxCut}.
\newblock \emph{ACM Transactions on Quantum Computing}, 4\penalty0
  (3):\penalty0 1--15, Sept 2023.
\newblock ISSN 2643-6809, 2643-6817.
\newblock \doi{10.1145/3584706}.
\newblock URL \url{http://arxiv.org/abs/2201.11785}.
\newblock arXiv:2201.11785 [quant-ph].

\bibitem[Augustino et~al.(2024)Augustino, Cain, Farhi, Gupta, Gutmann, Ranard,
  Tang, and Van~Kirk]{augustino2024strategies}
Brandon Augustino, Madelyn Cain, Edward Farhi, Swati Gupta, Sam Gutmann, Daniel
  Ranard, Eugene Tang, and Katherine Van~Kirk.
\newblock Strategies for running the {QAOA} at hundreds of qubits.
\newblock \emph{arXiv preprint arXiv:2410.03015}, 2024.

\bibitem[Hao et~al.(2024)Hao, He, Shaydulin, Larson, and Pistoia]{hao2024end}
Tianyi Hao, Zichang He, Ruslan Shaydulin, Jeffrey Larson, and Marco Pistoia.
\newblock End-to-end protocol for high-quality {QAOA} parameters with few
  shots.
\newblock \emph{arXiv preprint arXiv:2408.00557}, 2024.

\bibitem[Saffman et~al.(2010)Saffman, Walker, and
  M{\o}lmer]{saffman2010quantum}
Mark Saffman, Thad~G Walker, and Klaus M{\o}lmer.
\newblock Quantum information with {Rydberg} atoms.
\newblock \emph{Reviews of Modern Physics}, 82\penalty0 (3):\penalty0
  2313--2363, 2010.

\bibitem[Manetsch et~al.(2024)Manetsch, Nomura, Bataille, Leung, Lv, and
  Endres]{manetsch2024tweezer}
Hannah~J Manetsch, Gyohei Nomura, Elie Bataille, Kon~H Leung, Xudong Lv, and
  Manuel Endres.
\newblock A tweezer array with 6100 highly coherent atomic qubits.
\newblock \emph{arXiv preprint arXiv:2403.12021}, 2024.

\bibitem[Niu et~al.(2023)Niu, Hashim, Iancu, de~Jong, and
  Younis]{niu2023powerful}
Siyuan Niu, Akel Hashim, Costin Iancu, Wibe~Albert de~Jong, and Ed~Younis.
\newblock Powerful quantum circuit resizing with resource efficient synthesis.
\newblock \emph{arXiv preprint arXiv:2311.13107}, 2023.

\bibitem[Hashim et~al.(2024)Hashim, Yuan, Gokhale, Chen, Juenger, Fruitwala,
  Xu, Huang, Jiang, and Siddiqi]{hashim2024efficient}
Akel Hashim, Ming Yuan, Pranav Gokhale, Larry Chen, Christian Juenger, Neelay
  Fruitwala, Yilun Xu, Gang Huang, Liang Jiang, and Irfan Siddiqi.
\newblock Efficient generation of multi-partite entanglement between non-local
  superconducting qubits using classical feedback.
\newblock \emph{arXiv preprint arXiv:2403.18768}, 2024.

\bibitem[Delgado et~al.(2022)Delgado, Casares, dos Reis, Zini, Campos,
  Cruz-Hernández, Voigt, Lowe, Jahangiri, Martin-Delgado, Mueller, and
  Arrazola]{Delgado_2022}
Alain Delgado, Pablo A.~M. Casares, Roberto dos Reis, Modjtaba~Shokrian Zini,
  Roberto Campos, Norge Cruz-Hernández, Arne-Christian Voigt, Angus Lowe,
  Soran Jahangiri, M.~A. Martin-Delgado, Jonathan~E. Mueller, and Juan~Miguel
  Arrazola.
\newblock Simulating key properties of lithium-ion batteries with a
  fault-tolerant quantum computer.
\newblock \emph{Physical Review A}, 106\penalty0 (3), September 2022.
\newblock ISSN 2469-9934.
\newblock \doi{10.1103/physreva.106.032428}.
\newblock URL \url{http://dx.doi.org/10.1103/PhysRevA.106.032428}.

\bibitem[Shokrian~Zini et~al.(2023)Shokrian~Zini, Delgado, dos Reis,
  Moreno~Casares, Mueller, Voigt, and Arrazola]{ShokrianZini2023}
Modjtaba Shokrian~Zini, Alain Delgado, Roberto dos Reis, Pablo~Antonio
  Moreno~Casares, Jonathan~E. Mueller, Arne-Christian Voigt, and Juan~Miguel
  Arrazola.
\newblock Quantum simulation of battery materials using ionic pseudopotentials.
\newblock \emph{Quantum}, 7:\penalty0 1049, July 2023.
\newblock ISSN 2521-327X.
\newblock \doi{10.22331/q-2023-07-10-1049}.
\newblock URL \url{http://dx.doi.org/10.22331/q-2023-07-10-1049}.

\bibitem[Schuckert et~al.(2025)Schuckert, Kühn, Smith, Crane, and
  Girvin]{schuckert2025constrainedmanybodyphasesmathbbz2higgs}
Alexander Schuckert, Stefan Kühn, Kevin~C. Smith, Eleanor Crane, and Steven~M.
  Girvin.
\newblock Constrained many-body phases in a $\mathbb{Z}_2$-higgs lattice gauge
  theory.
\newblock 2025.
\newblock URL \url{https://arxiv.org/abs/2503.03828}.

\bibitem[Crane et~al.(2024)Crane, Smith, Tomesh, Eickbusch, Martyn, Kühn,
  Funcke, DeMarco, Chuang, Wiebe, Schuckert, and Girvin]{crane2024hybrid}
Eleanor Crane, Kevin~C. Smith, Teague Tomesh, Alec Eickbusch, John~M. Martyn,
  Stefan Kühn, Lena Funcke, Michael~Austin DeMarco, Isaac~L. Chuang, Nathan
  Wiebe, Alexander Schuckert, and Steven~M. Girvin.
\newblock Hybrid oscillator-qubit quantum processors: Simulating fermions,
  bosons, and gauge fields, 2024.

\bibitem[Schuckert et~al.(2024)Schuckert, Crane, Gorshkov, Hafezi, and
  Gullans]{schuckert2024fermion}
Alexander Schuckert, Eleanor Crane, Alexey~V. Gorshkov, Mohammad Hafezi, and
  Michael~J. Gullans.
\newblock Fermion-qubit fault-tolerant quantum computing, 2024.

\bibitem[Jordan et~al.(2012)Jordan, Lee, and
  Preskill]{Jordan_Lee_Preskill_2012}
Stephen~P. Jordan, Keith S.~M. Lee, and John Preskill.
\newblock Quantum algorithms for quantum field theories.
\newblock \emph{Science}, 336\penalty0 (6085):\penalty0 1130--1133, 2012.
\newblock \doi{10.1126/science.1217069}.

\bibitem[Shaw et~al.(2020)Shaw, Lougovski, Stryker, and Wiebe]{Shaw2020}
Alexander~F. Shaw, Pavel Lougovski, Jesse~R. Stryker, and Nathan Wiebe.
\newblock Quantum algorithms for simulating the lattice {Schwinger} model.
\newblock \emph{Quantum}, 4:\penalty0 306, August 2020.
\newblock ISSN 2521-327X.
\newblock \doi{10.22331/q-2020-08-10-306}.
\newblock URL \url{http://dx.doi.org/10.22331/q-2020-08-10-306}.

\bibitem[Sawaya et~al.(2020)Sawaya, Menke, Kyaw, Johri, Aspuru-Guzik, and
  Guerreschi]{Sawaya2020}
Nicolas P.~D. Sawaya, Tim Menke, Thi~Ha Kyaw, Sonika Johri, Alán Aspuru-Guzik,
  and Gian~Giacomo Guerreschi.
\newblock Resource-efficient digital quantum simulation of d-level systems for
  photonic, vibrational, and spin-s hamiltonians.
\newblock \emph{npj Quantum Information}, 6\penalty0 (1), June 2020.
\newblock ISSN 2056-6387.
\newblock \doi{10.1038/s41534-020-0278-0}.
\newblock URL \url{http://dx.doi.org/10.1038/s41534-020-0278-0}.

\bibitem[Liu et~al.(2024{\natexlab{a}})Liu, Singh, Smith, Crane, Martyn,
  Eickbusch, Schuckert, Li, Sinanan-Singh, Soley, Tsunoda, Chuang, Wiebe, and
  Girvin]{liu2024}
Yuan Liu, Shraddha Singh, Kevin~C. Smith, Eleanor Crane, John~M. Martyn, Alec
  Eickbusch, Alexander Schuckert, Richard~D. Li, Jasmine Sinanan-Singh,
  Micheline~B. Soley, Takahiro Tsunoda, Isaac~L. Chuang, Nathan Wiebe, and
  Steven~M. Girvin.
\newblock Hybrid oscillator-qubit quantum processors: Instruction set
  architectures, abstract machine models, and applications, 2024{\natexlab{a}}.

\bibitem[Häner et~al.(2018)Häner, Roetteler, and
  Svore]{häner2018optimizingquantumcircuitsarithmetic}
Thomas Häner, Martin Roetteler, and Krysta~M. Svore.
\newblock Optimizing quantum circuits for arithmetic, 2018.

\bibitem[Wang et~al.(2023)Wang, Frattini, Chapman, Puri, Girvin, Devoret, and
  Schoelkopf]{Wang2023}
Christopher~S. Wang, Nicholas~E. Frattini, Benjamin~J. Chapman, Shruti Puri,
  S.~M. Girvin, Michel~H. Devoret, and Robert~J. Schoelkopf.
\newblock Observation of wave-packet branching through an engineered conical
  intersection.
\newblock \emph{Phys. Rev. X}, 13:\penalty0 011008, Jan 2023.
\newblock \doi{10.1103/PhysRevX.13.011008}.
\newblock URL \url{https://link.aps.org/doi/10.1103/PhysRevX.13.011008}.

\bibitem[Whitlow et~al.(2023)Whitlow, Jia, Wang, Fang, Kim, and
  Brown]{whitlow2023}
Jacob Whitlow, Zhubing Jia, Ye~Wang, Chao Fang, Jungsang Kim, and Kenneth~R.
  Brown.
\newblock Quantum simulation of conical intersections using trapped ions.
\newblock \emph{Nature Chemistry}, 15\penalty0 (11):\penalty0 1509--1514,
  November 2023.
\newblock ISSN 1755-4349.
\newblock \doi{10.1038/s41557-023-01303-0}.

\bibitem[Shaw et~al.(2025)Shaw, Scholl, Finkelstein, Tsai, Choi, and
  Endres]{shaw2025}
Adam~L. Shaw, Pascal Scholl, Ran Finkelstein, Richard Bing-Shiun Tsai, Joonhee
  Choi, and Manuel Endres.
\newblock Erasure cooling, control, and hyperentanglement of motion in optical
  tweezers.
\newblock \emph{Science}, 388\penalty0 (6749):\penalty0 845--849, May 2025.
\newblock \doi{10.1126/science.adn2618}.

\bibitem[Stavenger et~al.(2022)Stavenger, Crane, Smith, Kang, Girvin, and
  Wiebe]{biskit}
Timothy~J Stavenger, Eleanor Crane, Kevin~C Smith, Christopher~T Kang, Steven~M
  Girvin, and Nathan Wiebe.
\newblock {C2QA -- Bosonic Qiskit}.
\newblock In \emph{2022 IEEE High Performance Extreme Computing Conference
  (HPEC)}, pages 1--8, 2022.
\newblock \doi{10.1109/HPEC55821.2022.9926318}.

\bibitem[Kang et~al.(2025)Kang, Soley, Crane, Girvin, and Wiebe]{Kang2025}
Christopher Kang, Micheline~B Soley, Eleanor Crane, Steven~M Girvin, and Nathan
  Wiebe.
\newblock Leveraging {Hamiltonian} simulation techniques to compile operations
  on bosonic devices.
\newblock \emph{Journal of Physics A: Mathematical and Theoretical},
  58\penalty0 (17):\penalty0 175301, April 2025.
\newblock ISSN 1751-8121.
\newblock \doi{10.1088/1751-8121/adb5df}.
\newblock URL \url{http://dx.doi.org/10.1088/1751-8121/adb5df}.

\bibitem[Decker et~al.(2025{\natexlab{a}})Decker, Gustafson, McKinney, Jones,
  Goetz, Li, Schuckert, Stein, Li, and Crane]{decker2025}
Ethan Decker, Erik Gustafson, Evan McKinney, Alex~K. Jones, Lucas Goetz, Ang
  Li, Alexander Schuckert, Samuel Stein, Gushu Li, and Eleanor Crane.
\newblock Symbolic {Hamiltonian} compiler for hybrid qubit-boson processors,
  2025{\natexlab{a}}.

\bibitem[Decker et~al.(2025{\natexlab{b}})Decker, Goetz, McKinney, Gustafson,
  Zhou, Liu, Jones, Li, Schuckert, Stein, Crane, and
  Li]{decker2025kernpilercompileroptimizationquantum}
Ethan Decker, Lucas Goetz, Evan McKinney, Erik Gustafson, Junyu Zhou, Yuhao
  Liu, Alex~K. Jones, Ang Li, Alexander Schuckert, Samuel Stein, Eleanor Crane,
  and Gushu Li.
\newblock Kernpiler: Compiler optimization for quantum {Hamiltonian} simulation
  with partial trotterization.
\newblock 2025{\natexlab{b}}.
\newblock URL \url{https://arxiv.org/abs/2504.07214}.

\bibitem[Chen et~al.(2025)Chen, Li, Guo, Chen, Li, Bierman, Huang, Zhou, Liu,
  and Zhang]{chen2025}
Zihan Chen, Jiakang Li, Minghao Guo, Henry Chen, Zirui Li, Joel Bierman, Yipeng
  Huang, Huiyang Zhou, Yuan Liu, and Eddy~Z. Zhang.
\newblock Genesis: A compiler framework for {Hamiltonian} simulation on hybrid
  {CV-DV} quantum computers, 2025.

\bibitem[Noh et~al.(2020)Noh, Girvin, and Jiang]{Noh_o2o}
Kyungjoo Noh, S.~M. Girvin, and Liang Jiang.
\newblock Encoding an oscillator into many oscillators.
\newblock \emph{Phys. Rev. Lett.}, 125:\penalty0 080503, Aug. 2020.
\newblock \doi{10.1103/PhysRevLett.125.080503}.
\newblock URL \url{https://link.aps.org/doi/10.1103/PhysRevLett.125.080503}.

\bibitem[Xu et~al.(2024)Xu, Wang, Vuillot, and
  Albert]{xu2024lettingtigercagebosonic}
Yijia Xu, Yixu Wang, Christophe Vuillot, and Victor~V. Albert.
\newblock Letting the tiger out of its cage: bosonic coding without
  concatenation, 2024.

\bibitem[Steckmann et~al.(2025)Steckmann, Luo, Wang, Muleady, Seif, Monroe,
  Gullans, Gorshkov, Katz, and
  Schuckert]{steckmann2025errormitigationshottoshotfluctuations}
Thomas Steckmann, De~Luo, Yu-Xin Wang, Sean~R. Muleady, Alireza Seif,
  Christopher Monroe, Michael~J. Gullans, Alexey~V. Gorshkov, Or~Katz, and
  Alexander Schuckert.
\newblock Error mitigation of shot-to-shot fluctuations in analog quantum
  simulators.
\newblock 2025.
\newblock URL \url{https://arxiv.org/abs/2506.16509}.

\bibitem[Verstraete and Cirac(2005)]{Verstraete_2005}
F~Verstraete and J~I Cirac.
\newblock Mapping local {Hamiltonians} of fermions to local {Hamiltonians} of
  spins.
\newblock \emph{Journal of Statistical Mechanics: Theory and Experiment},
  2005\penalty0 (09):\penalty0 P09012, Sept 2005.
\newblock \doi{10.1088/1742-5468/2005/09/P09012}.
\newblock URL \url{https://dx.doi.org/10.1088/1742-5468/2005/09/P09012}.

\bibitem[Derby et~al.(2021)Derby, Klassen, Bausch, and
  Cubitt]{Derby_compact_2021}
Charles Derby, Joel Klassen, Johannes Bausch, and Toby Cubitt.
\newblock Compact fermion to qubit mappings.
\newblock \emph{Phys. Rev. B}, 104:\penalty0 035118, Jul 2021.
\newblock \doi{10.1103/PhysRevB.104.035118}.
\newblock URL \url{https://link.aps.org/doi/10.1103/PhysRevB.104.035118}.

\bibitem[Luo and Cirac(2025)]{luo2025}
Maxine Luo and J.~Ignacio Cirac.
\newblock Efficient simulation of quantum chemistry problems in an enlarged
  basis set, 2025.

\bibitem[H{\'e}mery et~al.(2024)H{\'e}mery, Ghanem, Crane, Campbell, Dreiling,
  Figgatt, Foltz, Gaebler, Johansen, Mills, Moses, Pino, Ransford, Rowe,
  Siegfried, Stutz, Dreyer, Schuckert, and Nigmatullin]{hemery2024}
K{\'e}vin H{\'e}mery, Khaldoon Ghanem, Eleanor Crane, Sara~L. Campbell, Joan~M.
  Dreiling, Caroline Figgatt, Cameron Foltz, John~P. Gaebler, Jacob Johansen,
  Michael Mills, Steven~A. Moses, Juan~M. Pino, Anthony Ransford, Mary Rowe,
  Peter Siegfried, Russell~P. Stutz, Henrik Dreyer, Alexander Schuckert, and
  Ramil Nigmatullin.
\newblock Measuring the {{Loschmidt Amplitude}} for {{Finite-Energy
  Properties}} of the {{Fermi-Hubbard Model}} on an {{Ion-Trap Quantum
  Computer}}.
\newblock \emph{PRX Quantum}, 5\penalty0 (3):\penalty0 030323, August 2024.
\newblock \doi{10.1103/PRXQuantum.5.030323}.

\bibitem[Nigmatullin et~al.(2025)Nigmatullin, H{\'e}mery, Ghanem, Moses, Gresh,
  Siegfried, Mills, Gatterman, Hewitt, Granet, and Dreyer]{nigmatullin2025}
Ramil Nigmatullin, K{\'e}vin H{\'e}mery, Khaldoon Ghanem, Steven Moses, Dan
  Gresh, Peter Siegfried, Michael Mills, Thomas Gatterman, Nathan Hewitt,
  Etienne Granet, and Henrik Dreyer.
\newblock Experimental demonstration of breakeven for a compact fermionic
  encoding.
\newblock \emph{Nature Physics}, pages 1--7, June 2025.
\newblock ISSN 1745-2481.
\newblock \doi{10.1038/s41567-025-02931-8}.

\bibitem[Evered et~al.(2025)Evered, Kalinowski, Geim, Manovitz, Bluvstein, Li,
  Maskara, Zhou, Ebadi, Xu, Campo, Cain, Ostermann, Yelin, Sachdev, Greiner,
  Vuletić, and Lukin]{evered2025}
Simon~J. Evered, Marcin Kalinowski, Alexandra~A. Geim, Tom Manovitz, Dolev
  Bluvstein, Sophie~H. Li, Nishad Maskara, Hengyun Zhou, Sepehr Ebadi, Muqing
  Xu, Joseph Campo, Madelyn Cain, Stefan Ostermann, Susanne~F. Yelin, Subir
  Sachdev, Markus Greiner, Vladan Vuletić, and Mikhail~D. Lukin.
\newblock Probing topological matter and fermion dynamics on a neutral-atom
  quantum computer, 2025.

\bibitem[Kim et~al.(2023{\natexlab{a}})Kim, Eddins, Anand, Wei, {van den Berg},
  Rosenblatt, Nayfeh, Wu, Zaletel, Temme, and Kandala]{kim2023}
Youngseok Kim, Andrew Eddins, Sajant Anand, Ken~Xuan Wei, Ewout {van den Berg},
  Sami Rosenblatt, Hasan Nayfeh, Yantao Wu, Michael Zaletel, Kristan Temme, and
  Abhinav Kandala.
\newblock Evidence for the utility of quantum computing before fault tolerance.
\newblock \emph{Nature}, 618\penalty0 (7965):\penalty0 500--505, June
  2023{\natexlab{a}}.
\newblock ISSN 1476-4687.
\newblock \doi{10.1038/s41586-023-06096-3}.

\bibitem[Haghshenas et~al.(2025)Haghshenas, Chertkov, Mills, Kadow, Lin, Chen,
  Cade, Niesen, Begušić, Rudolph, Cirstoiu, Hemery, Keever, Lubasch, Granet,
  Baldwin, Bartolotta, Bohn, Cline, DeCross, Dreiling, Foltz, Francois,
  Gaebler, Gilbreth, Gray, Gresh, Hall, Hankin, Hansen, Hewitt, Hutson, Iqbal,
  Kotibhaskar, Lehman, Lucchetti, Madjarov, Mayer, Milne, Moses, Neyenhuis,
  Park, Ponsioen, Schecter, Siegfried, Stephen, Tiemann, Urmey, Walker, Potter,
  Hayes, Chan, Pollmann, Knap, Dreyer, and Foss-Feig]{haghshenas2025}
Reza Haghshenas, Eli Chertkov, Michael Mills, Wilhelm Kadow, Sheng-Hsuan Lin,
  Yi-Hsiang Chen, Chris Cade, Ido Niesen, Tomislav Begušić, Manuel~S.
  Rudolph, Cristina Cirstoiu, Kevin Hemery, Conor~Mc Keever, Michael Lubasch,
  Etienne Granet, Charles~H. Baldwin, John~P. Bartolotta, Matthew Bohn, Julia
  Cline, Matthew DeCross, Joan~M. Dreiling, Cameron Foltz, David Francois,
  John~P. Gaebler, Christopher~N. Gilbreth, Johnnie Gray, Dan Gresh, Alex Hall,
  Aaron Hankin, Azure Hansen, Nathan Hewitt, Ross~B. Hutson, Mohsin Iqbal,
  Nikhil Kotibhaskar, Elliot Lehman, Dominic Lucchetti, Ivaylo~S. Madjarov,
  Karl Mayer, Alistair~R. Milne, Steven~A. Moses, Brian Neyenhuis, Gunhee Park,
  Boris Ponsioen, Michael Schecter, Peter~E. Siegfried, David~T. Stephen,
  Bruce~G. Tiemann, Maxwell~D. Urmey, James Walker, Andrew~C. Potter, David
  Hayes, Garnet Kin-Lic Chan, Frank Pollmann, Michael Knap, Henrik Dreyer, and
  Michael Foss-Feig.
\newblock Digital quantum magnetism at the frontier of classical simulations,
  2025.

\bibitem[Yan et~al.(2022)Yan, Spar, Prichard, Chi, Wei,
  {Ibarra-Garc{\'i}a-Padilla}, Hazzard, and Bakr]{yan2022}
Zoe~Z. Yan, Benjamin~M. Spar, Max~L. Prichard, Sungjae Chi, Hao-Tian Wei,
  Eduardo {Ibarra-Garc{\'i}a-Padilla}, Kaden R.~A. Hazzard, and Waseem~S. Bakr.
\newblock Two-{{Dimensional Programmable Tweezer Arrays}} of {{Fermions}}.
\newblock \emph{Physical Review Letters}, 129\penalty0 (12):\penalty0 123201,
  September 2022.
\newblock \doi{10.1103/PhysRevLett.129.123201}.

\bibitem[{Gonz{\'a}lez-Cuadra} et~al.(2023){Gonz{\'a}lez-Cuadra}, Bluvstein,
  Kalinowski, Kaubruegger, Maskara, Naldesi, Zache, Kaufman, Lukin, Pichler,
  Vermersch, Ye, and Zoller]{gonzalez-cuadra2023}
D.~{Gonz{\'a}lez-Cuadra}, D.~Bluvstein, M.~Kalinowski, R.~Kaubruegger,
  N.~Maskara, P.~Naldesi, T.~V. Zache, A.~M. Kaufman, M.~D. Lukin, H.~Pichler,
  B.~Vermersch, Jun Ye, and P.~Zoller.
\newblock Fermionic quantum processing with programmable neutral atom arrays.
\newblock \emph{Proceedings of the National Academy of Sciences}, 120\penalty0
  (35):\penalty0 e2304294120, August 2023.
\newblock \doi{10.1073/pnas.2304294120}.

\bibitem[Wang et~al.(2022)Wang, Khatami, Fei, Wyrick, Namboodiri, Kashid,
  Rigosi, Bryant, and Silver]{wang2022a}
Xiqiao Wang, Ehsan Khatami, Fan Fei, Jonathan Wyrick, Pradeep Namboodiri,
  Ranjit Kashid, Albert~F. Rigosi, Garnett Bryant, and Richard Silver.
\newblock Experimental realization of an extended {{Fermi-Hubbard}} model using
  a {{2D}} lattice of dopant-based quantum dots.
\newblock \emph{Nature Communications}, 13\penalty0 (1):\penalty0 1--12,
  November 2022.
\newblock ISSN 2041-1723.
\newblock \doi{10.1038/s41467-022-34220-w}.

\bibitem[Rad et~al.(2024)Rad, Schuckert, Crane, Nambiar, Fei, Wyrick, Silver,
  Hafezi, Davoudi, and Gullans]{rad2024analogquantumsimulatorquantum}
Ali Rad, Alexander Schuckert, Eleanor Crane, Gautam Nambiar, Fan Fei, Jonathan
  Wyrick, Richard~M. Silver, Mohammad Hafezi, Zohreh Davoudi, and Michael~J.
  Gullans.
\newblock Analog quantum simulator of a quantum field theory with fermion-spin
  systems in silicon.
\newblock 2024.
\newblock URL \url{https://arxiv.org/abs/2407.03419}.

\bibitem[Ott et~al.(2024)Ott, {Gonz{\'a}lez-Cuadra}, Zache, Zoller, Kaufman,
  and Pichler]{ott2024}
Robert Ott, Daniel {Gonz{\'a}lez-Cuadra}, Torsten~V. Zache, Peter Zoller,
  Adam~M. Kaufman, and Hannes Pichler.
\newblock Error-corrected fermionic quantum processors with neutral atoms,
  2024.

\bibitem[O’Reilly and Olaya-Castro(2014)]{OReilly2014}
Edward~J. O’Reilly and Alexandra Olaya-Castro.
\newblock Non-classicality of the molecular vibrations assisting exciton energy
  transfer at room temperature.
\newblock \emph{Nature Communications}, 5:\penalty0 3012, January 2014.
\newblock ISSN 2041-1723.
\newblock \doi{10.1038/ncomms4012}.
\newblock URL \url{http://dx.doi.org/10.1038/ncomms4012}.

\bibitem[Liu et~al.(2021)Liu, Kolden, Krovi, Loureiro, Trivisa, and
  Childs]{Liu2021}
Jin-Peng Liu, Herman~{\O}ie Kolden, Hari~K. Krovi, Nuno~F. Loureiro,
  Konstantina Trivisa, and Andrew~M. Childs.
\newblock Efficient quantum algorithm for dissipative nonlinear differential
  equations.
\newblock \emph{Proceedings of the National Academy of Sciences}, 118\penalty0
  (35):\penalty0 e2026805118, August 2021.
\newblock \doi{10.1073/pnas.2026805118}.

\bibitem[Lu et~al.(2021)Lu, Ba{\~n}uls, and Cirac]{lu2021}
Sirui Lu, Mari~Carmen Ba{\~n}uls, and J.~Ignacio Cirac.
\newblock Algorithms for {{Quantum Simulation}} at {{Finite Energies}}.
\newblock \emph{PRX Quantum}, 2\penalty0 (2):\penalty0 020321, May 2021.
\newblock \doi{10.1103/PRXQuantum.2.020321}.

\bibitem[Schuckert et~al.(2023)Schuckert, Bohrdt, Crane, and
  Knap]{schuckert_2023}
Alexander Schuckert, Annabelle Bohrdt, Eleanor Crane, and Michael Knap.
\newblock Probing finite-temperature observables in quantum simulators of spin
  systems with short-time dynamics.
\newblock \emph{Phys. Rev. B}, 107:\penalty0 L140410, Apr 2023.
\newblock \doi{10.1103/PhysRevB.107.L140410}.
\newblock URL \url{https://link.aps.org/doi/10.1103/PhysRevB.107.L140410}.

\bibitem[Ghanem et~al.(2023)Ghanem, Schuckert, and Dreyer]{ghanem2023}
Khaldoon Ghanem, Alexander Schuckert, and Henrik Dreyer.
\newblock Robust {{Extraction}} of {{Thermal Observables}} from {{State
  Sampling}} and {{Real-Time Dynamics}} on {{Quantum Computers}}.
\newblock \emph{Quantum}, 7:\penalty0 1163, November 2023.
\newblock \doi{10.22331/q-2023-11-03-1163}.

\bibitem[Smith et~al.(2023)Smith, Crane, Wiebe, and
  Girvin]{PRXQuantum.4.020315}
Kevin~C. Smith, Eleanor Crane, Nathan Wiebe, and S.M. Girvin.
\newblock Deterministic constant-depth preparation of the aklt state on a
  quantum processor using fusion measurements.
\newblock \emph{PRX Quantum}, 4:\penalty0 020315, Apr 2023.
\newblock \doi{10.1103/PRXQuantum.4.020315}.
\newblock URL \url{https://link.aps.org/doi/10.1103/PRXQuantum.4.020315}.

\bibitem[Sajjan et~al.(2024)Sajjan, Singh, and Kais]{sajjan2024}
Manas Sajjan, Vinit Singh, and Sabre Kais.
\newblock Polynomially efficient quantum enabled variational {Monte {Carlo}}
  scheme for training neural-network quantum states for physico-chemical
  applications.
\newblock \emph{arXiv:2412.12398}, 2024.

\bibitem[Chen et~al.(2012)Chen, Gu, Liu, and Wen]{chen2012symmetry}
Xie Chen, Zheng-Cheng Gu, Zheng-Xin Liu, and Xiao-Gang Wen.
\newblock Symmetry-protected topological orders in interacting bosonic systems.
\newblock \emph{Science}, 338\penalty0 (6114):\penalty0 1604--1606, 2012.

\bibitem[Schuch et~al.(2011)Schuch, P{\'e}rez-Garc{\'\i}a, and
  Cirac]{schuch2011classifying}
Norbert Schuch, David P{\'e}rez-Garc{\'\i}a, and Ignacio Cirac.
\newblock Classifying quantum phases using matrix product states and projected
  entangled pair states.
\newblock \emph{Physical Review B—Condensed Matter and Materials Physics},
  84\penalty0 (16):\penalty0 165139, 2011.

\bibitem[Sajjan et~al.(2022)Sajjan, Alaeian, and Kais]{sajjan2022magnetic}
Manas Sajjan, Hadiseh Alaeian, and Sabre Kais.
\newblock Magnetic phases of spatially modulated spin-1 chains in {Rydberg}
  excitons: Classical and quantum simulations.
\newblock \emph{The Journal of Chemical Physics}, 157\penalty0 (22), 2022.

\bibitem[Lu et~al.(2019)Lu, Gao, and Duan]{PhysRevB.99.155136}
Sirui Lu, Xun Gao, and L.-M. Duan.
\newblock Efficient representation of topologically ordered states with
  restricted {Boltzmann} machines.
\newblock \emph{Phys. Rev. B}, 99:\penalty0 155136, April 2019.
\newblock \doi{10.1103/PhysRevB.99.155136}.
\newblock URL \url{https://link.aps.org/doi/10.1103/PhysRevB.99.155136}.

\bibitem[Gao and Duan(2017)]{gao2017efficient}
Xun Gao and Lu-Ming Duan.
\newblock Efficient representation of quantum many-body states with deep neural
  networks.
\newblock \emph{Nature Communications}, 8\penalty0 (1):\penalty0 662, 2017.

\bibitem[Luo et~al.(2022)Luo, Chen, Carrasquilla, and
  Clark]{luo2022autoregressive}
Di~Luo, Zhuo Chen, Juan Carrasquilla, and Bryan~K Clark.
\newblock Autoregressive neural network for simulating open quantum systems via
  a probabilistic formulation.
\newblock \emph{Physical Review Letters}, 128\penalty0 (9):\penalty0 090501,
  2022.

\bibitem[Yoshioka and Hamazaki(2019)]{PhysRevB.99.214306}
Nobuyuki Yoshioka and Ryusuke Hamazaki.
\newblock Constructing neural stationary states for open quantum many-body
  systems.
\newblock \emph{Phys. Rev. B}, 99:\penalty0 214306, Jun 2019.
\newblock \doi{10.1103/PhysRevB.99.214306}.
\newblock URL \url{https://link.aps.org/doi/10.1103/PhysRevB.99.214306}.

\bibitem[Wang et~al.(2020)Wang, Navarrete-Benlloch, and
  Cai]{PhysRevLett.125.115301}
Zijian Wang, Carlos Navarrete-Benlloch, and Zi~Cai.
\newblock Pattern formation and exotic order in driven-dissipative
  {Bose--{Hubbard}} systems.
\newblock \emph{Phys. Rev. Lett.}, 125:\penalty0 115301, Sep 2020.
\newblock \doi{10.1103/PhysRevLett.125.115301}.
\newblock URL \url{https://link.aps.org/doi/10.1103/PhysRevLett.125.115301}.

\bibitem[Sun et~al.(2021)Sun, Motta, Tazhigulov, Tan, Chan, and
  Minnich]{PRXQuantum.2.010317}
Shi-Ning Sun, Mario Motta, Ruslan~N. Tazhigulov, Adrian~T.K. Tan, Garnet
  Kin-Lic Chan, and Austin~J. Minnich.
\newblock Quantum computation of finite-temperature static and dynamical
  properties of spin systems using quantum imaginary time evolution.
\newblock \emph{PRX Quantum}, 2:\penalty0 010317, Feb. 2021.
\newblock \doi{10.1103/PRXQuantum.2.010317}.
\newblock URL \url{https://link.aps.org/doi/10.1103/PRXQuantum.2.010317}.

\bibitem[Tamascelli et~al.(2019)Tamascelli, Smirne, Lim, Huelga, and
  Plenio]{PhysRevLett.123.090402}
D.~Tamascelli, A.~Smirne, J.~Lim, S.~F. Huelga, and M.~B. Plenio.
\newblock Efficient simulation of finite-temperature open quantum systems.
\newblock \emph{Phys. Rev. Lett.}, 123:\penalty0 090402, Aug. 2019.
\newblock \doi{10.1103/PhysRevLett.123.090402}.
\newblock URL \url{https://link.aps.org/doi/10.1103/PhysRevLett.123.090402}.

\bibitem[Powers et~al.(2023)Powers, Bassman~Oftelie, Camps, and
  de~Jong]{powers2023exploring}
Connor Powers, Lindsay Bassman~Oftelie, Daan Camps, and Wibe~A de~Jong.
\newblock Exploring finite temperature properties of materials with quantum
  computers.
\newblock \emph{Scientific Reports}, 13\penalty0 (1):\penalty0 1986, 2023.

\bibitem[Motta et~al.(2020)Motta, Sun, Tan, O’Rourke, Ye, Minnich, Brandao,
  and Chan]{motta2020determining}
Mario Motta, Chong Sun, Adrian~TK Tan, Matthew~J O’Rourke, Erika Ye, Austin~J
  Minnich, Fernando~GSL Brandao, and Garnet Kin-Lic Chan.
\newblock Determining eigenstates and thermal states on a quantum computer
  using quantum imaginary time evolution.
\newblock \emph{Nature Physics}, 16\penalty0 (2):\penalty0 205--210, 2020.

\bibitem[Irikura and Saito(2020)]{PhysRevResearch.2.013284}
Naoki Irikura and Hiroki Saito.
\newblock Neural-network quantum states at finite temperature.
\newblock \emph{Phys. Rev. Res.}, 2:\penalty0 013284, Mar 2020.
\newblock \doi{10.1103/PhysRevResearch.2.013284}.
\newblock URL \url{https://link.aps.org/doi/10.1103/PhysRevResearch.2.013284}.

\bibitem[Kemper et~al.(2016)Kemper, Vafek, Betts, Balakirev, Hardy, Liang,
  Bonn, and Boebinger]{kemper2016thermodynamic}
JB~Kemper, O~Vafek, JB~Betts, FF~Balakirev, WN~Hardy, Ruixing Liang, DA~Bonn,
  and GS~Boebinger.
\newblock Thermodynamic signature of a magnetic-field-driven phase transition
  within the superconducting state of an underdoped cuprate.
\newblock \emph{Nature Physics}, 12\penalty0 (1):\penalty0 47--51, 2016.

\bibitem[Heyl et~al.(2013)Heyl, Polkovnikov, and Kehrein]{heyl2013dynamical}
Markus Heyl, Anatoli Polkovnikov, and Stefan Kehrein.
\newblock Dynamical quantum phase transitions in the transverse-field {Ising}
  model.
\newblock \emph{Physical Review Letters}, 110\penalty0 (13):\penalty0 135704,
  2013.

\bibitem[Kloss et~al.(2020)Kloss, Reichman, and Bar~Lev]{kloss2020studying}
Benedikt Kloss, David Reichman, and Yevgeny Bar~Lev.
\newblock Studying dynamics in two-dimensional quantum lattices using tree
  tensor network states.
\newblock \emph{SciPost Physics}, 9\penalty0 (5):\penalty0 070, 2020.

\bibitem[Mendoza-Arenas(2022)]{mendoza2022dynamical}
Juan~Jos{\'e} Mendoza-Arenas.
\newblock Dynamical quantum phase transitions in the one-dimensional extended
  {Fermi--{Hubbard}} model.
\newblock \emph{Journal of Statistical Mechanics: Theory and Experiment},
  2022\penalty0 (4):\penalty0 043101, 2022.

\bibitem[Budich and Heyl(2016)]{budich2016dynamical}
Jan~Carl Budich and Markus Heyl.
\newblock Dynamical topological order parameters far from equilibrium.
\newblock \emph{Physical Review B}, 93\penalty0 (8):\penalty0 085416, 2016.

\bibitem[Hryniuk and Szyma{\'n}ska(2024)]{hryniuk2024tensor}
Dawid~A Hryniuk and Marzena~H Szyma{\'n}ska.
\newblock Tensor-network-based variational {Monte {Carlo}} approach to the
  non-equilibrium steady state of open quantum systems.
\newblock \emph{arXiv preprint arXiv:2405.12044}, 2024.

\bibitem[Mori et~al.(2018)Mori, Ikeda, Kaminishi, and
  Ueda]{mori2018thermalization}
Takashi Mori, Tatsuhiko~N Ikeda, Eriko Kaminishi, and Masahito Ueda.
\newblock Thermalization and prethermalization in isolated quantum systems: a
  theoretical overview.
\newblock \emph{Journal of Physics B: Atomic, Molecular and Optical Physics},
  51\penalty0 (11):\penalty0 112001, 2018.

\bibitem[Cazalilla et~al.(2012)Cazalilla, Iucci, and
  Chung]{cazalilla2012thermalization}
Miguel~A Cazalilla, Anibal Iucci, and Ming-Chiang Chung.
\newblock Thermalization and quantum correlations in exactly solvable models.
\newblock \emph{Physical Review E -- Statistical, Nonlinear, and Soft Matter
  Physics}, 85\penalty0 (1):\penalty0 011133, 2012.

\bibitem[Sajjan et~al.(2023)Sajjan, Singh, Selvarajan, and
  Kais]{PhysRevResearch.5.013146}
Manas Sajjan, Vinit Singh, Raja Selvarajan, and Sabre Kais.
\newblock Imaginary components of out-of-time-order correlator and information
  scrambling for navigating the learning landscape of a quantum machine
  learning model.
\newblock \emph{Phys. Rev. Res.}, 5:\penalty0 013146, Feb 2023.
\newblock \doi{10.1103/PhysRevResearch.5.013146}.
\newblock URL \url{https://link.aps.org/doi/10.1103/PhysRevResearch.5.013146}.

\bibitem[Cerezo et~al.(2021)Cerezo, Arrasmith, Babbush, Benjamin, Endo, Fujii,
  McClean, Mitarai, Yuan, Cincio, and Coles]{Cerezo_2021}
M.~Cerezo, Andrew Arrasmith, Ryan Babbush, Simon~C. Benjamin, Suguru Endo,
  Keisuke Fujii, Jarrod~R. McClean, Kosuke Mitarai, Xiao Yuan, Lukasz Cincio,
  and Patrick~J. Coles.
\newblock Variational quantum algorithms.
\newblock \emph{Nature Reviews Physics}, 3\penalty0 (9):\penalty0 625--644,
  August 2021.
\newblock ISSN 2522-5820.
\newblock \doi{10.1038/s42254-021-00348-9}.
\newblock URL \url{http://dx.doi.org/10.1038/s42254-021-00348-9}.

\bibitem[Grimsley et~al.(2019)Grimsley, Economou, Barnes, and
  Mayhall]{Grimsley_2019}
Harper~R. Grimsley, Sophia~E. Economou, Edwin Barnes, and Nicholas~J. Mayhall.
\newblock An adaptive variational algorithm for exact molecular simulations on
  a quantum computer.
\newblock \emph{Nature Communications}, 10\penalty0 (1), July 2019.
\newblock ISSN 2041-1723.
\newblock \doi{10.1038/s41467-019-10988-2}.
\newblock URL \url{http://dx.doi.org/10.1038/s41467-019-10988-2}.

\bibitem[Shaydulin et~al.(2019)Shaydulin, Safro, and Larson]{Shaydulin_2019}
Ruslan Shaydulin, Ilya Safro, and Jeffrey Larson.
\newblock Multistart methods for quantum approximate optimization.
\newblock In \emph{2019 IEEE High Performance Extreme Computing Conference
  (HPEC)}. IEEE, Sept 2019.
\newblock \doi{10.1109/hpec.2019.8916288}.
\newblock URL \url{http://dx.doi.org/10.1109/HPEC.2019.8916288}.

\bibitem[Bravyi et~al.(2024)Bravyi, Cross, Gambetta, Maslov, Rall, and
  Yoder]{Bravyi_2024}
Sergey Bravyi, Andrew~W. Cross, Jay~M. Gambetta, Dmitri Maslov, Patrick Rall,
  and Theodore~J. Yoder.
\newblock High-threshold and low-overhead fault-tolerant quantum memory.
\newblock \emph{Nature}, 627\penalty0 (8005):\penalty0 778--782, March 2024.
\newblock ISSN 1476-4687.
\newblock \doi{10.1038/s41586-024-07107-7}.
\newblock URL \url{http://dx.doi.org/10.1038/s41586-024-07107-7}.

\bibitem[Poole et~al.(2024)Poole, Graham, Perlin, Otten, and
  Saffman]{Poole_2024}
C.~Poole, T.~M. Graham, M.~A. Perlin, M.~Otten, and M.~Saffman.
\newblock Architecture for fast implementation of {qLDPC} codes with optimized
  {Rydberg} gates, 2024.
\newblock URL \url{https://arxiv.org/abs/2404.18809}.

\bibitem[Giles and Selinger(2013)]{Giles_2013}
Brett Giles and Peter Selinger.
\newblock Exact synthesis of multiqubit {Clifford+T} circuits.
\newblock \emph{Physical Review A}, 87\penalty0 (3), March 2013.
\newblock ISSN 1094-1622.
\newblock \doi{10.1103/physreva.87.032332}.
\newblock URL \url{http://dx.doi.org/10.1103/PhysRevA.87.032332}.

\bibitem[Liu et~al.(2024{\natexlab{b}})Liu, Gonzales, Huang, Saleem, and
  Hovland]{Liu_2024}
Ji~Liu, Alvin Gonzales, Benchen Huang, Zain~Hamid Saleem, and Paul Hovland.
\newblock {QuCLEAR: Clifford} extraction and absorption for significant
  reduction in quantum circuit size, 2024{\natexlab{b}}.
\newblock URL \url{https://arxiv.org/abs/2408.13316}.

\bibitem[Larocca et~al.(2025)Larocca, Thanasilp, Wang, Sharma, Biamonte, Coles,
  Cincio, McClean, Holmes, and Cerezo]{larocca2025barren}
Mart{\'\i}n Larocca, Supanut Thanasilp, Samson Wang, Kunal Sharma, Jacob
  Biamonte, Patrick~J Coles, Lukasz Cincio, Jarrod~R McClean, Zo{\"e} Holmes,
  and M~Cerezo.
\newblock Barren plateaus in variational quantum computing.
\newblock \emph{Nature Reviews Physics}, pages 1--16, 2025.

\bibitem[Ryan-Anderson et~al.(2021)Ryan-Anderson, Bohnet, Lee, Gresh, Hankin,
  Gaebler, Francois, Chernoguzov, Lucchetti, Brown,
  et~al.]{ryan2021realization}
Ciaran Ryan-Anderson, Justin~G Bohnet, Kenny Lee, Daniel Gresh, Aaron Hankin,
  JP~Gaebler, David Francois, Alexander Chernoguzov, Dominic Lucchetti,
  Natalie~C Brown, et~al.
\newblock Realization of real-time fault-tolerant quantum error correction.
\newblock \emph{Physical Review X}, 11\penalty0 (4):\penalty0 041058, 2021.

\bibitem[{Google Quantum AI}(2023)]{google2023suppressing}
{Google Quantum AI}.
\newblock Suppressing quantum errors by scaling a surface code logical qubit.
\newblock \emph{Nature}, 614\penalty0 (7949):\penalty0 676--681, 2023.

\bibitem[Sivak et~al.(2023)Sivak, Eickbusch, Royer, Singh, Tsioutsios, Ganjam,
  Miano, Brock, Ding, Frunzio, et~al.]{sivak2023real}
Volodymyr~V Sivak, Alec Eickbusch, Baptiste Royer, Shraddha Singh, Ioannis
  Tsioutsios, Suhas Ganjam, Alessandro Miano, BL~Brock, AZ~Ding, Luigi Frunzio,
  et~al.
\newblock Real-time quantum error correction beyond break-even.
\newblock \emph{Nature}, 616\penalty0 (7955):\penalty0 50--55, 2023.

\bibitem[Ryan-Anderson et~al.(2024)Ryan-Anderson, Brown, Baldwin, Dreiling,
  Foltz, Gaebler, Gatterman, Hewitt, Holliman, Horst, et~al.]{ryan2024high}
C~Ryan-Anderson, NC~Brown, CH~Baldwin, JM~Dreiling, C~Foltz, JP~Gaebler,
  TM~Gatterman, N~Hewitt, C~Holliman, CV~Horst, et~al.
\newblock High-fidelity teleportation of a logical qubit using transversal
  gates and lattice surgery.
\newblock \emph{Science}, 385\penalty0 (6715):\penalty0 1327--1331, 2024.

\bibitem[Bluvstein et~al.(2024)Bluvstein, Evered, Geim, Li, Zhou, Manovitz,
  Ebadi, Cain, Kalinowski, Hangleiter, et~al.]{bluvstein2024logical}
Dolev Bluvstein, Simon~J Evered, Alexandra~A Geim, Sophie~H Li, Hengyun Zhou,
  Tom Manovitz, Sepehr Ebadi, Madelyn Cain, Marcin Kalinowski, Dominik
  Hangleiter, et~al.
\newblock Logical quantum processor based on reconfigurable atom arrays.
\newblock \emph{Nature}, 626\penalty0 (7997):\penalty0 58--65, 2024.

\bibitem[Gupta et~al.(2024)Gupta, Sundaresan, Alexander, Wood, Merkel, Healy,
  Hillenbrand, Jochym-O’Connor, Wootton, Yoder, et~al.]{gupta2024encoding}
Riddhi~S Gupta, Neereja Sundaresan, Thomas Alexander, Christopher~J Wood,
  Seth~T Merkel, Michael~B Healy, Marius Hillenbrand, Tomas Jochym-O’Connor,
  James~R Wootton, Theodore~J Yoder, et~al.
\newblock Encoding a magic state with beyond break-even fidelity.
\newblock \emph{Nature}, 625\penalty0 (7994):\penalty0 259--263, 2024.

\bibitem[{Lanyon} et~al.(2010){Lanyon}, {Whitfield}, {Gillett}, {Goggin},
  {Almeida}, {Kassal}, {Biamonte}, {Mohseni}, {Powell}, {Barbieri},
  {Aspuru-Guzik}, and {White}]{lanyon2010towards}
B.~P. {Lanyon}, J.~D. {Whitfield}, G.~G. {Gillett}, M.~E. {Goggin}, M.~P.
  {Almeida}, I.~{Kassal}, J.~D. {Biamonte}, M.~{Mohseni}, B.~J. {Powell},
  M.~{Barbieri}, A.~{Aspuru-Guzik}, and A.~G. {White}.
\newblock Towards quantum chemistry on a quantum computer.
\newblock \emph{Nature Chemistry}, 2\penalty0 (2):\penalty0 106--111, 2010.

\bibitem[Peruzzo et~al.(2014)Peruzzo, McClean, Shadbolt, Yung, Zhou, Love,
  Aspuru-Guzik, and O?brien]{peruzzo2014variational}
Alberto Peruzzo, Jarrod McClean, Peter Shadbolt, Man-Hong Yung, Xiao-Qi Zhou,
  Peter~J Love, Al{\'a}n Aspuru-Guzik, and Jeremy~L O?brien.
\newblock A variational eigenvalue solver on a photonic quantum processor.
\newblock \emph{Nature Communications}, 5:\penalty0 4213, 2014.

\bibitem[Wang et~al.(2015)Wang, Dolde, Biamonte, Babbush, Bergholm, Yang,
  Jakobi, Neumann, Aspuru-Guzik, Whitfield, et~al.]{wang2015quantum}
Ya~Wang, Florian Dolde, Jacob Biamonte, Ryan Babbush, Ville Bergholm, Sen Yang,
  Ingmar Jakobi, Philipp Neumann, Al{\'a}n Aspuru-Guzik, James~D Whitfield,
  et~al.
\newblock Quantum simulation of helium hydride cation in a solid-state spin
  register.
\newblock \emph{ACS Nano}, 9\penalty0 (8):\penalty0 7769--7774, 2015.

\bibitem[Kandala et~al.(2017)Kandala, Mezzacapo, Temme, Takita, Brink, Chow,
  and Gambetta]{kandala2017hardware}
Abhinav Kandala, Antonio Mezzacapo, Kristan Temme, Maika Takita, Markus Brink,
  Jerry~M Chow, and Jay~M Gambetta.
\newblock Hardware-efficient variational quantum eigensolver for small
  molecules and quantum magnets.
\newblock \emph{Nature}, 549\penalty0 (7671):\penalty0 242--246, 2017.

\bibitem[and others(2016)]{o2016scalable}
and others.
\newblock Scalable quantum simulation of molecular energies.
\newblock \emph{Physical Review X}, 6\penalty0 (3):\penalty0 031007, 2016.

\bibitem[Hempel et~al.(2018)Hempel, Maier, Romero, McClean, Monz, Shen,
  Jurcevic, Lanyon, Love, Babbush, et~al.]{hempel2018quantum}
Cornelius Hempel, Christine Maier, Jonathan Romero, Jarrod McClean, Thomas
  Monz, Heng Shen, Petar Jurcevic, Ben~P Lanyon, Peter Love, Ryan Babbush,
  et~al.
\newblock Quantum chemistry calculations on a trapped-ion quantum simulator.
\newblock \emph{Physical Review X}, 8\penalty0 (3):\penalty0 031022, 2018.

\bibitem[Dumitrescu et~al.(2018)Dumitrescu, McCaskey, Hagen, Jansen, Morris,
  Papenbrock, Pooser, Dean, and Lougovski]{dumitrescu2018cloud}
Eugene~F Dumitrescu, Alex~J McCaskey, Gaute Hagen, Gustav~R Jansen, Titus~D
  Morris, Thomas Papenbrock, Raphael~C Pooser, David~Jarvis Dean, and Pavel
  Lougovski.
\newblock Cloud quantum computing of an atomic nucleus.
\newblock \emph{Physical Review Letters}, 120\penalty0 (21):\penalty0 210501,
  2018.

\bibitem[Quantum et~al.(2020)Quantum, Collaborators*†, Arute, Arya, Babbush,
  Bacon, Bardin, Barends, Boixo, Broughton, Buckley, et~al.]{google2020hartree}
Google~AI Quantum, Collaborators*†, Frank Arute, Kunal Arya, Ryan Babbush,
  Dave Bacon, Joseph~C Bardin, Rami Barends, Sergio Boixo, Michael Broughton,
  Bob~B Buckley, et~al.
\newblock {Hartree--{Fock}} on a superconducting qubit quantum computer.
\newblock \emph{Science}, 369\penalty0 (6507):\penalty0 1084--1089, 2020.

\bibitem[Kreshchuk et~al.(2021)Kreshchuk, Jia, Kirby, Goldstein, Vary, and
  Love]{kreshchuk2021light}
Michael Kreshchuk, Shaoyang Jia, William~M Kirby, Gary Goldstein, James~P Vary,
  and Peter~J Love.
\newblock Light-front field theory on current quantum computers.
\newblock \emph{Entropy}, 23\penalty0 (5):\penalty0 597, 2021.

\bibitem[Preskill(2018)]{preskill2018quantum}
John Preskill.
\newblock Quantum computing in the {NISQ} era and beyond.
\newblock \emph{Quantum}, 2:\penalty0 79, 2018.

\bibitem[Arute et~al.(2019)Arute, Arya, Babbush, Bacon, Bardin, Barends,
  Biswas, Boixo, Brandao, Buell, et~al.]{arute2019quantum}
Frank Arute, Kunal Arya, Ryan Babbush, Dave Bacon, Joseph~C Bardin, Rami
  Barends, Rupak Biswas, Sergio Boixo, Fernando~GSL Brandao, David~A Buell,
  et~al.
\newblock Quantum supremacy using a programmable superconducting processor.
\newblock \emph{Nature}, 574\penalty0 (7779):\penalty0 505--510, 2019.

\bibitem[Turing et~al.(1936)]{turing1936computable}
Alan~Mathison Turing et~al.
\newblock On computable numbers, with an application to the
  {Entscheidungsproblem}.
\newblock \emph{J. of Math}, 58\penalty0 (345-363):\penalty0 5, 1936.

\bibitem[Church(1936)]{church1936unsolvable}
Alonzo Church.
\newblock An unsolvable problem of elementary number theory.
\newblock \emph{American Journal of Mathematics}, 58\penalty0 (2):\penalty0
  345--363, 1936.

\bibitem[Robledo-Moreno et~al.(2024)Robledo-Moreno, Motta, Haas, Javadi-Abhari,
  Jurcevic, Kirby, Martiel, Sharma, Sharma, Shirakawa, Sitdikov, Sun, Sung,
  Takita, Tran, Yunoki, and Mezzacapo]{robledo2024chemistry}
Javier Robledo-Moreno, Mario Motta, Holger Haas, Ali Javadi-Abhari, Petar
  Jurcevic, William Kirby, Simon Martiel, Kunal Sharma, Sandeep Sharma,
  Tomonori Shirakawa, Iskandar Sitdikov, Rong-Yang Sun, Kevin~J. Sung, Maika
  Takita, Minh~C. Tran, Seiji Yunoki, and Antonio Mezzacapo.
\newblock Chemistry beyond exact solutions on a quantum-centric supercomputer,
  2024.

\bibitem[Farrell et~al.(2024)Farrell, Illa, Ciavarella, and
  Savage]{farrell2024scalable}
Roland~C Farrell, Marc Illa, Anthony~N Ciavarella, and Martin~J Savage.
\newblock Scalable circuits for preparing ground states on digital quantum
  computers: The {Schwinger} model vacuum on 100 qubits.
\newblock \emph{PRX Quantum}, 5\penalty0 (2):\penalty0 020315, 2024.

\bibitem[Lee et~al.(2023)Lee, Lee, Zhai, Tong, Dalzell, Kumar, Helms, Gray,
  Cui, Liu, et~al.]{lee2023evaluating}
Seunghoon Lee, Joonho Lee, Huanchen Zhai, Yu~Tong, Alexander~M Dalzell,
  Ashutosh Kumar, Phillip Helms, Johnnie Gray, Zhi-Hao Cui, Wenyuan Liu, et~al.
\newblock Evaluating the evidence for exponential quantum advantage in
  ground-state quantum chemistry.
\newblock \emph{Nature Communications}, 14\penalty0 (1):\penalty0 1952, 2023.

\bibitem[Boixo et~al.(2018)Boixo, Isakov, Smelyanskiy, Babbush, Ding, Jiang,
  Bremner, Martinis, and Neven]{boixo2018characterizing}
Sergio Boixo, Sergei~V Isakov, Vadim~N Smelyanskiy, Ryan Babbush, Nan Ding,
  Zhang Jiang, Michael~J Bremner, John~M Martinis, and Hartmut Neven.
\newblock Characterizing quantum supremacy in near-term devices.
\newblock \emph{Nature Physics}, 14\penalty0 (6):\penalty0 595--600, 2018.

\bibitem[Aharonov et~al.(2023)Aharonov, Gao, Landau, Liu, and
  Vazirani]{aharonov2023polynomial}
Dorit Aharonov, Xun Gao, Zeph Landau, Yunchao Liu, and Umesh Vazirani.
\newblock A polynomial-time classical algorithm for noisy random circuit
  sampling.
\newblock In \emph{Proceedings of the 55th Annual ACM Symposium on Theory of
  Computing}, pages 945--957, 2023.

\bibitem[Pednault et~al.(2019)Pednault, Gunnels, Nannicini, Horesh, and
  Wisnieff]{pednault2019leveraging}
Edwin Pednault, John~A Gunnels, Giacomo Nannicini, Lior Horesh, and Robert
  Wisnieff.
\newblock Leveraging secondary storage to simulate deep 54-qubit sycamore
  circuits.
\newblock \emph{arXiv preprint arXiv:1910.09534}, 2019.

\bibitem[Oh et~al.(2023)Oh, Jiang, and Fefferman]{oh2023spoofing}
Changhun Oh, Liang Jiang, and Bill Fefferman.
\newblock Spoofing cross-entropy measure in boson sampling.
\newblock \emph{Physical Review Letters}, 131\penalty0 (1):\penalty0 010401,
  2023.

\bibitem[Kim et~al.(2023{\natexlab{b}})Kim, Eddins, Anand, Wei, Van Den~Berg,
  Rosenblatt, Nayfeh, Wu, Zaletel, Temme, et~al.]{kim2023evidence}
Youngseok Kim, Andrew Eddins, Sajant Anand, Ken~Xuan Wei, Ewout Van Den~Berg,
  Sami Rosenblatt, Hasan Nayfeh, Yantao Wu, Michael Zaletel, Kristan Temme,
  et~al.
\newblock Evidence for the utility of quantum computing before fault tolerance.
\newblock \emph{Nature}, 618\penalty0 (7965):\penalty0 500--505,
  2023{\natexlab{b}}.

\bibitem[Tindall et~al.(2024)Tindall, Fishman, Stoudenmire, and
  Sels]{tindall2024efficient}
Joseph Tindall, Matthew Fishman, E~Miles Stoudenmire, and Dries Sels.
\newblock Efficient tensor network simulation of {IBM}’s eagle kicked {Ising}
  experiment.
\newblock \emph{PRX Quantum}, 5\penalty0 (1):\penalty0 010308, 2024.

\bibitem[Begu{\v{s}}i{\'c} et~al.(2024)Begu{\v{s}}i{\'c}, Gray, and
  Chan]{beguvsic2024fast}
Tomislav Begu{\v{s}}i{\'c}, Johnnie Gray, and Garnet Kin-Lic Chan.
\newblock Fast and converged classical simulations of evidence for the utility
  of quantum computing before fault tolerance.
\newblock \emph{Science Advances}, 10\penalty0 (3):\penalty0 eadk4321, 2024.

\bibitem[Liao et~al.(2023)Liao, Wang, Zhou, Zhang, and
  Xiang]{liao2023simulation}
Hai-Jun Liao, Kang Wang, Zong-Sheng Zhou, Pan Zhang, and Tao Xiang.
\newblock Simulation of {IBM}'s kicked {Ising} experiment with projected
  entangled pair operator.
\newblock \emph{arXiv preprint arXiv:2308.03082}, 2023.

\bibitem[King et~al.(2023)King, Raymond, Lanting, Harris, Zucca, Altomare,
  Berkley, Boothby, Ejtemaee, Enderud, et~al.]{king2023quantum}
Andrew~D King, Jack Raymond, Trevor Lanting, Richard Harris, Alex Zucca, Fabio
  Altomare, Andrew~J Berkley, Kelly Boothby, Sara Ejtemaee, Colin Enderud,
  et~al.
\newblock Quantum critical dynamics in a 5,000-qubit programmable spin glass.
\newblock \emph{Nature}, pages 1--6, 2023.

\bibitem[King et~al.(2025{\natexlab{a}})King, Nocera, Rams, Dziarmaga,
  Wiersema, Bernoudy, Raymond, Kaushal, Heinsdorf, Harris,
  et~al.]{king2025beyond}
Andrew~D King, Alberto Nocera, Marek~M Rams, Jacek Dziarmaga, Roeland Wiersema,
  William Bernoudy, Jack Raymond, Nitin Kaushal, Niclas Heinsdorf, Richard
  Harris, et~al.
\newblock Beyond-classical computation in quantum simulation.
\newblock \emph{Science}, 388\penalty0 (6743):\penalty0 199--204,
  2025{\natexlab{a}}.

\bibitem[Tindall et~al.(2025)Tindall, Mello, Fishman, Stoudenmire, and
  Sels]{tindall2025dynamics}
Joseph Tindall, Antonio Mello, Matt Fishman, Miles Stoudenmire, and Dries Sels.
\newblock Dynamics of disordered quantum systems with two-and three-dimensional
  tensor networks.
\newblock \emph{arXiv preprint arXiv:2503.05693}, 2025.

\bibitem[Mauron and Carleo(2025)]{mauron2025challenging}
Linda Mauron and Giuseppe Carleo.
\newblock Challenging the quantum advantage frontier with large-scale classical
  simulations of annealing dynamics.
\newblock \emph{arXiv preprint arXiv:2503.08247}, 2025.

\bibitem[Park et~al.(2025)Park, Gray, and Chan]{park2025simulating}
Gunhee Park, Johnnie Gray, and Garnet~Kin Chan.
\newblock Simulating quantum dynamics in two-dimensional lattices with tensor
  network influence functional belief propagation.
\newblock \emph{arXiv preprint arXiv:2504.07344}, 2025.

\bibitem[King et~al.(2025{\natexlab{b}})King, Nocera, Rams, Dziarmaga, Raymond,
  Kaushal, Sandvik, Alvarez, Carrasquilla, Franz, et~al.]{king2025comment}
Andrew~D King, Alberto Nocera, Marek~M Rams, Jacek Dziarmaga, Jack Raymond,
  Nitin Kaushal, Anders~W Sandvik, Gonzalo Alvarez, Juan Carrasquilla, Marcel
  Franz, et~al.
\newblock Comment on: "dynamics of disordered quantum systems with two-and
  three-dimensional tensor networks" arxiv: 2503.05693.
\newblock \emph{arXiv preprint arXiv:2504.06283}, 2025{\natexlab{b}}.

\bibitem[Baker(2016)]{baker20161}
Monya Baker.
\newblock 1,500 scientists lift the lid on reproducibility, 2016.

\bibitem[Chambers and Tzavella(2022)]{chambers2022past}
Christopher~D Chambers and Loukia Tzavella.
\newblock The past, present and future of registered reports.
\newblock \emph{Nature Human Behaviour}, 6\penalty0 (1):\penalty0 29--42, 2022.

\bibitem[III()]{CCCBD}
Russell D.~Johnson III.
\newblock {NIST} {C}omputational {C}hemistry {C}omparison and {B}enchmark
  {D}atabase, {NIST} {S}tandard {R}eference {D}atabase {N}umber 101, {R}elease
  22, {M}ay 2022.
\newblock URL \url{http://cccbdb.nist.gov/}.

\bibitem[Buro and B{\"u}ning(1992)]{buro1992report}
Michael Buro and H~Kleine B{\"u}ning.
\newblock \emph{Report on a {SAT} Competition}.
\newblock Fachbereich Math.-Informatik, Univ. Gesamthochschule Paderborn,
  Germany, 1992.

\bibitem[Simon et~al.(2005)Simon, Le~Berre, and Hirsch]{simon2005sat2002}
Laurent Simon, Daniel Le~Berre, and Edward~A Hirsch.
\newblock {The SAT2002} competition.
\newblock \emph{Annals of Mathematics and Artificial Intelligence},
  43:\penalty0 307--342, 2005.

\bibitem[Le~Berre and Simon(2003)]{le2003essentials}
Daniel Le~Berre and Laurent Simon.
\newblock The essentials of the {SAT} 2003 competition.
\newblock In \emph{International Conference on Theory and Applications of
  Satisfiability Testing}, pages 452--467. Springer, 2003.

\bibitem[Kautz et~al.(2004)Kautz, Selman, and McAllester]{kautz2004walksat}
Henry Kautz, Bart Selman, and David McAllester.
\newblock Walksat in the {2004 SAT} competition.
\newblock In \emph{Proceedings of the International Conference on Theory and
  Applications of Satisfiability Testing}, 2004.

\bibitem[J{\"a}rvisalo et~al.(2012)J{\"a}rvisalo, Le~Berre, Roussel, and
  Simon]{jarvisalo2012international}
Matti J{\"a}rvisalo, Daniel Le~Berre, Olivier Roussel, and Laurent Simon.
\newblock The international {SAT} solver competitions.
\newblock \emph{Ai Magazine}, 33\penalty0 (1):\penalty0 89--92, 2012.

\bibitem[Balyo et~al.(2017)Balyo, Heule, and Jarvisalo]{balyo2017sat}
Tom{\'a}s Balyo, Marijn Heule, and Matti Jarvisalo.
\newblock {SAT} competition 2016: Recent developments.
\newblock In \emph{Proceedings of the AAAI Conference on Artificial
  Intelligence}, volume~31, 2017.

\bibitem[Heule et~al.(2019)Heule, J{\"a}rvisalo, and Suda]{heule2019sat}
Marijn~JH Heule, Matti J{\"a}rvisalo, and Martin Suda.
\newblock {SAT} competition 2018.
\newblock \emph{Journal on Satisfiability, Boolean Modelling and Computation},
  11\penalty0 (1):\penalty0 133--154, 2019.

\bibitem[Kochemazov(2020)]{kochemazov2020improving}
Stepan Kochemazov.
\newblock Improving implementation of {SAT} competitions 2017--2019 winners.
\newblock In \emph{International Conference on Theory and Applications of
  Satisfiability Testing}, pages 139--148. Springer, 2020.

\bibitem[Huberman et~al.(1997)Huberman, Lukose, and
  Hogg]{huberman1997economics}
Bernardo~A Huberman, Rajan~M Lukose, and Tad Hogg.
\newblock An economics approach to hard computational problems.
\newblock \emph{Science}, 275\penalty0 (5296):\penalty0 51--54, 1997.

\bibitem[Maymin(2011)]{efficientNP}
Philip~Z. Maymin.
\newblock Markets are efficient if and only if {P}={NP}.
\newblock \emph{Algorithmic Finance}, 1.1:\penalty0 1--11, 2011.

\bibitem[Pennock(2001)]{problemmarkets}
David~M. Pennock.
\newblock {NP} markets, or how to get everyone else to solve your intractable
  problems.
\newblock In \emph{Workshop on Economic Agents, Models, and Mechanisms at the
  17th International Joint Conference on Artificial Intelligence (IJCAI)},
  2001.

\end{thebibliography}

\end{document}